\documentclass[final,3p,times,twocolumn]{elsarticle}
\usepackage{amsmath,amssymb,amsbsy,graphicx,color,xcolor,epsfig}
\usepackage{color,colortbl}

\definecolor{darkishgreen}{RGB}{39,203,22}
\definecolor{LightCyan}{rgb}{0.88,1,1}
\definecolor{Gray}{gray}{0.9}
\definecolor{lightRed}{RGB}{230,170,150}
\definecolor{modRed}{RGB}{230,82,90}
\definecolor{strongRed}{RGB}{230,6,6} 

\usepackage{multirow}
\usepackage{fullpage}
\usepackage{tabularx}
\usepackage{booktabs} 
\usepackage{orcidlink} 

\biboptions{sort&compress}

\journal{Physics Letters B}

\begin{document}

\begin{frontmatter}


\title{
\hfill DESY 23-036 \\[5mm]
{\bf Branching Fraction of the Decay $B^+ \to \boldsymbol\pi^+ \boldsymbol\tau^+ \boldsymbol\tau^-$ and} \\ 
{\bf Lepton Flavor Universality Test via the Ratio $R_{\boldsymbol\pi} (\boldsymbol\tau/\boldsymbol\mu)$}   
%  Branching Fraction of the Decay $B^+ \to \pi^+ \tau^+ \tau^-$ and \\ 
%  Lepton Flavor Universality Test via the Ratio $R_\pi (\tau/\mu)$   
}

\author[label1]{Ahmed Ali\,\orcidlink{0000-0002-1939-1545}} 

\affiliation[label1]{organization= {Deutsches Elektronen-Synchrotron DESY},
            addressline={Notkestr. 85},
           city={22607 Hamburg},
           country={Germany}}

\author{Alexander Ya. Parkhomenko\,\orcidlink{0000-0003-3773-5820}} 

\author{Irina M. Parnova\,\orcidlink{0000-0002-3205-0569}}
            
\begin{abstract}
Among (semi)leptonic rare $B$-decays induced by the $b \to d$ flavor changing neutral current,  
the decay $B^+ \to \pi^+ \mu^+ \mu^-$ is the only one observed so far experimentally. 
Related decays involving the $e^+e^-$ and $\tau^+ \tau^-$ pairs are the targets for the ongoing 
experiments at the LHC, in particular LHCb, and Belle II. The muonic and electronic semileptonic 
decays have almost identical branching fractions in the Standard Model (SM). However, the tauonic 
decay $B^+ \to \pi^+ \tau^+ \tau^-$ differs from the other two due to the higher reaction threshold 
which lies slightly below the $\psi (2S)$-resonance. We present calculations of the ditauon ($\tau^+ \tau^-$)
invariant-mass distribution and the branching fraction ${\rm Br} (B^+ \to \pi^+ \tau^+ \tau^-) $ 
in the SM based on the Effective Electroweak Hamiltonian approach, taking into account also 
the so-called long-distance contributions. The largest theoretical uncertainty in the 
short-distance part of the decay rates is due to the $B \to \pi$ form factors, which we quantify 
using three popular parametrizations. The long-distance contribution can be minimized by a cut 
on the ditauon mass $m_{\tau^+ \tau^-} >  M_{\psi (2S)}$. Once available, the branching fractions 
in the tauonic and muonic (and electronic) modes provide stringent test of the lepton flavor 
universality in the $b \to d$ transitions. We illustrate this by calculating the ratio 
$R_\pi (\tau/\mu) \equiv {\rm Br} (B^+ \to \pi^+ \tau^+ \tau^-)/{\rm Br} (B^+ \to \pi^+ \mu^+ \mu^-) $
in the SM for the total and binned ratios of the branching fractions.  

\end{abstract}


\end{frontmatter}

\section{Introduction}
\label{sec:Introduction}

Rare bottom-hadron decays induced by the quark-level Flavor Changing Neutral Current (FCNC) 
transitions $b \to s$ and $b \to d$ are of special interest as they  allow us to test the SM 
precisely and search for possible deviations from the Standard Model (SM). FCNC processes 
in the SM are governed by the GIM mechanism~\cite{Glashow:1970gm}, which allows such transitions  
only through higher-order electroweak (loop) diagrams. In particular, semileptonic rare decays 
are a very useful tool for testing the Lepton Flavor Universality (LFU), a linchpin of the 
electroweak sector of the SM. Semileptonic decays due to the $b \to s$ currents such as 
$B^\pm \to K^{(*)\pm} \mu^+ \mu^-$, $B^0 \to K^{(*)0} \mu^+ \mu^-$, and $B_s^0 \to \phi \mu^+ \mu^-$ 
and their electronic counterparts, while suppressed by the loops, are favored 
by the quark mixing Cabibbo-Kobayashi-Maskawa (CKM) matrix~\cite{Cabibbo:1963yz,Kobayashi:1973fv}.
Hence, there is plenty of data available on their branching fractions and decay characteristics, 
such as the lepton-pair invariant-mass and angular distributions~\cite{%
Aaij:2014pli,Aaij:2014tfa,Aaij:2015esa,Aaij:2015oid,Aaij:2016flj,LHCb:2021zwz,LHCb:2021xxq}. 
Some of these measurements were found to be not in accord with the SM-based predictions, triggering 
searches for better models incorporating physics beyond the Standard Model (BSM)~\cite{%
Celis:2017doq,Buttazzo:2017ixm,Aebischer:2019mlg,Alasfar:2020mne,Isidori:2021tzd,Ciuchini:2022wbq}. 
 
An important issue in these decays is the interference between the short (perturbative)- and 
long (non-perturbative)-distance  contributions. The standard experimental procedure is 
to exclude the dilepton invariant-mass squared ($q^2$) spectrum close to the $J/\psi$- and 
$\psi (2S)$-resonances, and extract the short-distance part of the spectrum from the rest. 
% There are also several broad vector charmonium resonances with masses above the open charm threshold.
Measurements of the phase difference between the short- and long-distance amplitudes 
in the $B^+ \to K^+ \mu^+ \mu^-$ decay has been undertaken by the LHCb collaboration based 
on data collected in 2011 and 2012~\cite{LHCb:2016due}. Their analysis shows that the phases 
of the $J/\psi$- and $\psi (2S)$-mesons are important near the resonance masses, 
due to their small decay widths, but their influence on the dilepton invariant mass spectrum 
in other regions is small. In addition, the branching fractions of the higher charmonium states: 
$\psi (3770)$, $\psi (4040)$, $\psi (4160)$, and $\psi (4415)$, were measured. This analysis 
is potentially helpful for studies of other semileptonic decays, $B^+ \to \pi^+ \ell^+ \ell^-$, 
in particular. For the $B \to \pi e^+ e^-$ and $B \to \pi \mu^+ \mu^-$, also light vector mesons, 
$\rho^0$, $\omega$ and $\phi$, give sizable contributions to the lepton invariant-mass distribution 
around $q^2 \sim 1$~GeV$^2$. 

Dedicated searches of possible LFU-violations in rare decays due to the $b \to s$ currents 
have been undertaken by the LHCb collaboration~\cite{LHCb:2017avl,LHCb:2019hip,LHCb:2021trn} 
in terms of the ratios $R_{K^{(*)}} (\mu/e)\equiv
{\rm Br} (B \to K^{(*)} \mu^+ \mu^-)/{\rm Br} (B \to K^{(*)} e^+ e^-)$, measured 
in selected bins in the dilepton invariant mass squared. This data hinted at LFU-violation, 
typically reaching three standard deviation from the SM. Similar analysis by the Belle 
collaboration~\cite{BELLE:2019xld,Belle:2019oag}, on the other hand, yielding
$R_{K^{(*)}} (\mu/e) = 1.03^{+0.28}_{-0.24} \pm 0.01$ for $q^2 \in (1.0,\, 6.0)$ GeV$^2$, 
while consistent with the SM is less conclusive due to larger experimental errors. However, 
recent measurements of the ratios $R_{K^{(*)}} (\mu/e)$ in the low- and central-$q^2$ parts 
of the spectrum by the LHCb collaboration~\cite{LHCb:2022qnv,LHCb:2022zom} are found in almost 
perfect agreement with the SM predictions~\cite{Hiller:2003js,Bordone:2016gaq}. This data, 
based on 9~fb$^{-1}$ integrated luminosity, with improved understanding of the background 
and tighter electron particle identification, supersedes the earlier LHCb data.

There have also been persistent indications over almost a decade of LFU-violation 
in the charged current (CC) semileptonic transitions $B \to D^{(*)} \ell \nu_\ell$, 
comparing the light $(\ell = e,\, \mu)$ and $\tau$-lepton modes via the ratios
$R_{D^{(*)}} \equiv {\rm Br} (B \to D^{(*)} \tau \nu_\tau)/{\rm Br} (B \to D^{(*)} \ell \nu_\ell)$~\cite{% 
LHCb:2015gmp,BaBar:2013mob,Belle:2016dyj,LHCb:2017rln}. However, the latest analysis of~$R_{D^*}$ 
by the LHCb collaboration~\cite{LHCb-LHCC,LHCb:2023cjr}, yielding 
$R_{D^{*-}} = {\rm Br} (B^0 \to D^{*-} \tau^+ \nu_\tau)/{\rm Br}(B^0 \to D^{*-} \mu^+ \nu_\mu) = 
0.247 \pm 0.015\, ({\rm stat}) \pm 0.015\, ({\rm syst}) \pm 0.012\, ({\rm ext})$,
is in good agreement with the SM-based estimate $R_{D^*} = 0.254 \pm 0.005$~\cite{HFLAV:2022}. 
Likewise, the single best-measurement of~$R_D$, namely 
$R_D = 0.307 \pm 0.037\, ({\rm stat}) \pm 0.016\, ({\rm syst}) $ by the Belle 
collaboration~\cite{Belle:2019rba}, is in good agreement with the corresponding ratio in the SM, 
$R_D = 0.298 \pm 0.004$~\cite{HFLAV:2022}. Thus, the long-standing anomalies in~$R_D$ 
and~$R_{D^*}$ in CC decays have receded, thanks to precise data. We also mention that 
the 2.6 standard-deviation departure from the LFU observed by the LEP experiments 
in the branching fractions of the $W^\pm \to \ell ^\pm \nu_\ell$ decays, namely 
$R_{\tau/(e + \mu)}^{\rm LEP} \equiv 2 {\rm Br} (W^\pm \to \tau^\pm \nu_\tau)/
[{\rm Br} (W^\pm \to e^\pm \nu_e) + {\rm Br} (W^\pm \to \mu^\pm \nu_\mu)) = 1.066 \pm 0.025$~\cite{% 
ALEPH:2013dgf,ParticleDataGroup:2022pth}, has now been brought in line with the SM expectations 
$R_{\tau/(e + \mu)}^{\rm SM} = 0.9996$~\cite{Denner:1991kt,Kniehl:2000rb}, 
by precise experiments in proton-proton collisions at the LHC with 
$R_{\tau/(e + \mu)}^{\rm CMS}= 1.002 \pm 0.019$~\cite{CMS:2022mhs}.
Measurements by the ATLAS collaboration~\cite{ATLAS:2020xea}, 
$R_{\mu/e }^{\rm ATLAS}= 1.003 \pm 0.010$ and $R_{\tau/\mu}^{\rm ATLAS}= 0.992 \pm 0.013$,  
are likewise in excellent agreement with the LFU hypothesis. One concludes that 
there is no experimental evidence of the LFU-violation in charged-current processes.
 
Data on the FCNC semileptonic $b \to d$ transitions is rather sparse. 
For decays induced by the $b \to d \ell^+ \ell^-$ transition, where $\ell = e,\, \mu,\, \tau$, 
the $B^+ \to \pi^+ \mu^+ \mu^-$ decay is so far the only mode observed in the $B$-meson sector, 
first reported by the LHCb collaboration in 2012~\cite{LHCb:2012de} and analyzed in detail 
in 2015~\cite{Aaij:2015nea}. The measured dimuon invariant mass distribution in the 
$B^+ \to \pi^+ \mu^+ \mu^-$ decay is in good agreement with theoretical predictions 
in the SM~\cite{Ali:2013zfa,Faustov:2014zva,Hou:2014dza} in almost all regions of the spectrum, 
except the lowest $q^2$-part. In this region, experimental data significantly exceeds 
theoretical predictions based on the short-distance contribution~\cite{Aaij:2015nea}. 
Taking into account the sub-leading weak annihilation~(WA) and long-distance~(LD) contributions, 
however, gives better agreement between theoretical predictions and experimental 
data~\cite{Hambrock:2015wka,Ali:2020tjy,Parnova:2022qfb}. We also note the evidences 
for the $B^0 \to \pi^+ \pi^- \mu^+ \mu^-$ decay with a significance of~$4.8\sigma$~\cite{LHCb:2014yov},
the $B_s^0 \to K^{*0} \mu^+ \mu^-$ decay at $3.4\sigma$~\cite{LHCb:2018rym}, reported 
by the LHCb collaboration, and the observation of the $\Lambda^0_b \to p \pi^- \mu^+ \mu^-$ 
decay in the bottom-baryon sector by the same collaboration~\cite{LHCb:2017lpt},
all of them are mediated by the $b \to d \ell^+ \ell^-$ transition.  
A model-independent analysis of the $|\Delta b| = |\Delta d| = 1$ processes to test the SM 
and probe flavor patterns of new physics was undertaken in~\cite{Bause:2022rrs}. 
Constraints on Wilson coefficients are obtained from global fits from data on exclusive 
$B^+ \to \pi^+ \mu^+ \mu^-$, $B_s \to \bar K^{*0} \mu^+ \mu^-$, $B^0 \to \mu^+ \mu^-$, 
and inclusive radiative $B \to X_d \gamma$ decays. Being consistent with the SM,  
these fits leave a sizable room for new physics.
The ratio $R_\pi (\mu/e)$, involving the branching ratios of $B^\pm \to \pi^\pm \ell^+ \ell^-$ 
for $\ell^\pm =e^\pm, \mu^\pm$, has been studied theoretically at great length to search 
for the LFU-violation~\cite{Bordone:2021olx} in the semileptonic $b \to d$ sector. 
At present, however, there is no data available on the ratio $R_\pi (\mu/e)$.

Precision tests of LFU involving the decays $b \to (s,\, d)\, \tau^+ \tau^-$ remain to be undertaken.
Compared to the $b \to (s,\, d)\, e^+ e^-$ and $b \to (s,\, d)\, \mu^+ \mu^-$ modes, they 
have a reduced phase space, in addition to the experimental difficulty of reconstructing 
the $\tau^\pm$-leptons. These modes will be targeted at the LHCb and Belle II experiments. 
Theoretically, they have the advantage of being relatively free of the LD-contributions and 
the form factors involved in the SD-piece can eventually be calculated precisely on the lattice. 
In the $b \to s$ sector, the decays $B \to K^{(*)} \tau^+ \tau^-$ have recently been studied theoretically 
from the BSM physics point of view~\cite{Capdevila:2017iqn,Crivellin:2018yvo,Alguero:2022wkd,SinghChundawat:2022ldm}. 
Currently only weak experimental upper limits on these decays are available, with
${\rm Br} (B^+ \to K^+ \tau^+ \tau^-) < 2.25 \times 10^{-3}$ by the BaBar collaboration~\cite{BaBar:2016wgb} 
and ${\rm Br} (B^0 \to K^{\ast 0} \tau^+ \tau^-) < 2.0 \times 10^{-3}$ by Belle~\cite{Belle:2021ndr},   
both obtained at $90\%$~CL.

In the $b \to d \tau^+ \tau^-$ sector, the main decays of interest are $B^+ \to \pi^+ \tau^+ \tau^-$, 
$B^0 \to \pi^+ \pi^-  \tau^+ \tau^-$, $B^0 \to \rho^0 \tau^+ \tau^-$ and  $B^+ \to \rho^+ \tau^+ \tau^-$. 
In this paper, we present a theoretical analysis of the $B^+ \to \pi^+ \tau^+ \tau^-$ decay 
in the SM and work out the ditauon invariant-mass distribution and decay width for three popular 
parametrizations of the $B \to \pi$ transition form factors~\cite{Boyd:1995sq,Bourrely:2008za,Leljak:2021vte}. 
The LD-contributions are calculated using the available data on the decay chain 
$B \to \pi V \to \pi \ell^+ \ell^-$~\cite{ParticleDataGroup:2022pth}, which can, however, 
be greatly reduced by imposing a cut on the dilepton invariant mass, 
$m_{\ell^+ \ell^-} > M_{\psi(2S)}$. We also estimate the ratio of the tauonic-to-muonic 
branching fractions, $R_\pi (\tau/\mu)$, which holds also for the ratio $R_\pi (\tau/e)$ in the SM.
Their measurements will test the LFU-violations involving all three charged leptons 
in the FCNC $b \to d$ sector.

\section{Effective Hamiltonian for the $b \to d \ell^+ \ell^-$ decays in the SM}
\label{sec:Theory}

Our analysis is carried out in the Effective Electroweak Hamiltonians approach~\cite{Buchalla:1995vs,Chetyrkin:1996vx}, 
where the SM heavy degrees of freedom $(W^\pm, Z^0, t)$ are absent. This effective theory also 
does not contain photons and gluons with energies exceeding the mass of the $b$-quark,~$m_b$, 
which represents the largest energy scale of the theory. Photons and gluons with lower energies 
are included using the QED and QCD Lagrangians. 
Rare semileptonic decays of the $B$-mesons involving the $b \to s$ and $b \to d$ FCNC transitions 
are calculated in this framework, of which the $b \to d$ part has the form:
%~\cite{Chetyrkin:1996vx}:
% 
\begin{eqnarray}
&& \hspace*{-13mm} 
{\cal H}^{b \to d}_{\rm weak} =  
\frac{4 G_F}{\sqrt 2} \Bigg \{ V_{ud} V_{ub}^* \left [ 
C_1 (\mu)\, {\cal P}^{(u)}_1 (\mu) + C_2 (\mu)\, {\cal P}^{(u)}_2 (\mu)  
\right ] 
\nonumber \\ 
&& \hspace*{-5mm} 
+ V_{cd} V_{cb}^* \left [ 
C_1 (\mu)\, {\cal P}^{(c)}_1 (\mu) + C_2 (\mu)\, {\cal P}^{(c)}_2 (\mu)  
\right ] 
\label{eq:Heff} \\ 
&& \hspace*{-5mm} 
- V_{td} V_{tb}^* \sum\limits_{j=3}^{10} C_j (\mu)\, {\cal P}_j (\mu)  
\Bigg \} + {\rm h.\, c.} ,
\nonumber 
\end{eqnarray}
where~$G_F$ is the Fermi constant, $V_{q_1 q_2}$ are the CKM matrix elements 
satisfying the unitary condition $V_{ud} V_{ub}^* + V_{cd} V_{cb}^* + V_{td} V_{tb}^* = 0$  
which can be used to eliminate one of their products, and $C_j (\mu)$ are Wilson coefficients 
determined at the scale~$\mu$. For the operators ${\cal P}_j (\mu)$, the following basis 
is chosen~\cite{Chetyrkin:1996vx,Bobeth:1999mk}:
\begin{equation}
{\cal P}^{(p)}_1 = 
(\bar d \gamma_\mu L T^A p)\, (\bar p \gamma^\mu L T^A b), 
\label{eq:P-1}
\end{equation} 
\begin{equation}
{\cal P}^{(p)}_2 = 
(\bar d \gamma_\mu L p)\, (\bar p \gamma^\mu L b),
\label{eq:P-2}
\end{equation}
\begin{equation}
{\cal P}_3 = 
(\bar d \gamma_\mu L b) \sum_q (\bar q \gamma^\mu q), 
\label{eq:P-3}
\end{equation}
\begin{equation}
{\cal P}_4 = 
(\bar d \gamma_\mu L T^A b) \sum_q (\bar q \gamma^\mu T^A q), 
\label{eq:P-4}
\end{equation}
\begin{equation}
{\cal P}_5 = 
(\bar d \gamma_\mu \gamma_\nu \gamma_\rho L b) \sum_q (\bar q \gamma^\mu \gamma^\nu \gamma^\rho q), 
\label{eq:P-5}
\end{equation}
\begin{equation}
{\cal P}_6 = 
(\bar d \gamma_\mu \gamma_\nu \gamma_\rho L T^A b) \sum_q (\bar q \gamma^\mu \gamma^\nu \gamma^\rho T^A q), 
\label{eq:P-6}
\end{equation}
\begin{equation}
\hspace{-3mm}{\cal P}_{7\gamma} = \frac{e}{16 \pi^2} \left [ 
\bar d \sigma^{\mu\nu} (m_b R + m_d L) b 
\right ] F_{\mu\nu}, 
\label{eq:P-7}
\end{equation}
\begin{equation}
{\cal P}_{8g} = \frac{g_{\rm st}}{16 \pi^2} \left [ 
\bar d \sigma^{\mu\nu} (m_b R + m_d L) T^A b
\right ] G^A_{\mu\nu},
\label{eq:P-8}
\end{equation}
\begin{equation}
{\cal P}_{9\ell} = \frac{\alpha_{\rm em}}{2 \pi}
(\bar d \gamma_\mu L b) \sum_\ell (\bar \ell \gamma^\mu \ell),
\label{eq:P-9}
\end{equation}
\begin{equation} 
{\cal P}_{10\ell} = \frac{\alpha_{\rm em}}{2 \pi}
(\bar d \gamma_\mu L b) \sum_\ell (\bar \ell \gamma^\mu \gamma^5 \ell),
\label{eq:P-10}
\end{equation}
where $p = u,\, c$ is the quark flavor, 
$T^A$ $(A = 1,\, \ldots,\, 8)$ are the generators of the color $SU (3)_C$-group, 
$L, R = \left ( 1 \mp \gamma_5 \right) /2$ are the left- and right-handed fermionic projectors,
$F_{\mu\nu}$ and $G^A_{\mu\nu}$ are the electromagnetic and gluon field strength tensors, respectively,
$m_b$ and $m_d$ are the $b$- and $d$-quark masses of which the $d$-quark mass is neglected,
$\sigma_{\mu \nu} = i \left(\gamma_\mu \gamma_\nu - \gamma_\nu \gamma_\mu \right)/2$, and 
$\alpha_{\rm em} = e^2/(4\pi)$ is the fine structure constant. The summation over~$q$ and~$\ell$ 
denotes sums over all quarks (except the $t$-quark) and charged leptons, respectively.
The Wilson coefficients $C_j (\mu)$, which depend on the renormalization scale $\mu$, are 
calculated at the matching scale $\mu_W \sim m_W$, where $m_W$ is the $W$-boson mass, 
as a perturbative expansion in the strong coupling constant $\alpha_s (\mu_W)$~\cite{Bobeth:1999mk}:
\begin{equation}
C_j (\mu_W) = \sum_{k = 0}^\infty 
\left [ \frac{\alpha_s (\mu_W)}{4 \pi} \right ]^k C_j^{(k)} (\mu_W) ,  
\end{equation}
which are evolved to a lower scale $\mu_b \sim m_b$ using
the anomalous dimensions of the above operators. They have been calculated  to the
next-next-leading-log (NNLL) accuracy~\cite{Bobeth:1999mk}:
\begin{equation} 
\gamma_i = 
  \frac{\alpha_s (\mu_W)}{4 \pi} \, \gamma^{(0)}_i  
+ \left ( \frac{\alpha_s (\mu_W)}{4 \pi} \right )^2 \gamma^{(1)}_i  
+ \left ( \frac{\alpha_s (\mu_W)}{4 \pi} \right )^3 \gamma^{(2)}_i 
+ \ldots    
\label{eq:AD-expansion}
\end{equation}
Numerical values of the Wilson coefficient, calculated to NLL accuracy,  are presented in Table~\ref{table:Wilson-coeff-mb}, 
where one can see that the Wilson coefficients of the QCD penguin operators, $C_j (m_b)$  
with $j = 3,\, 4,\, 5,\, 6$, have much smaller values than the others.

\begin{table}[tb]
% \caption{Numerical values of the Wilson coefficients at the scale $\mu_b = m_b = 4.8$~GeV.} 
\caption{Wilson coefficients at the scale $\mu_b = m_b = 4.8$~GeV.} 
\label{table:Wilson-coeff-mb}
% 
% {\small
\begin{center} 
\begin{tabular}{|cc|cc|}\hline 
$C_1 (m_b)$ & $-0.146$ &  $C_2 (m_b)$ & $1.056$ \\ 
$C_3 (m_b)$ & $0.011$ & $C_4 (m_b)$ & $-0.033$ \\
$C_5 (m_b)$ & $0.010$ & $C_6 (m_b)$ & $-0.039$ \\ 
$C_{7\gamma} (m_b)$ & $-0.317$ & $C_{8g} (m_b)$ & $0.149$ \\ 
$C_{9\ell} (m_b)$ & $4.15$ & $C_{10\ell} (m_b)$ & $-4.26$ \\   
\hline    
\end{tabular}
\end{center}
% }
\end{table}

\begin{figure}[tb]
\centerline{
\includegraphics[width=0.24\textwidth]{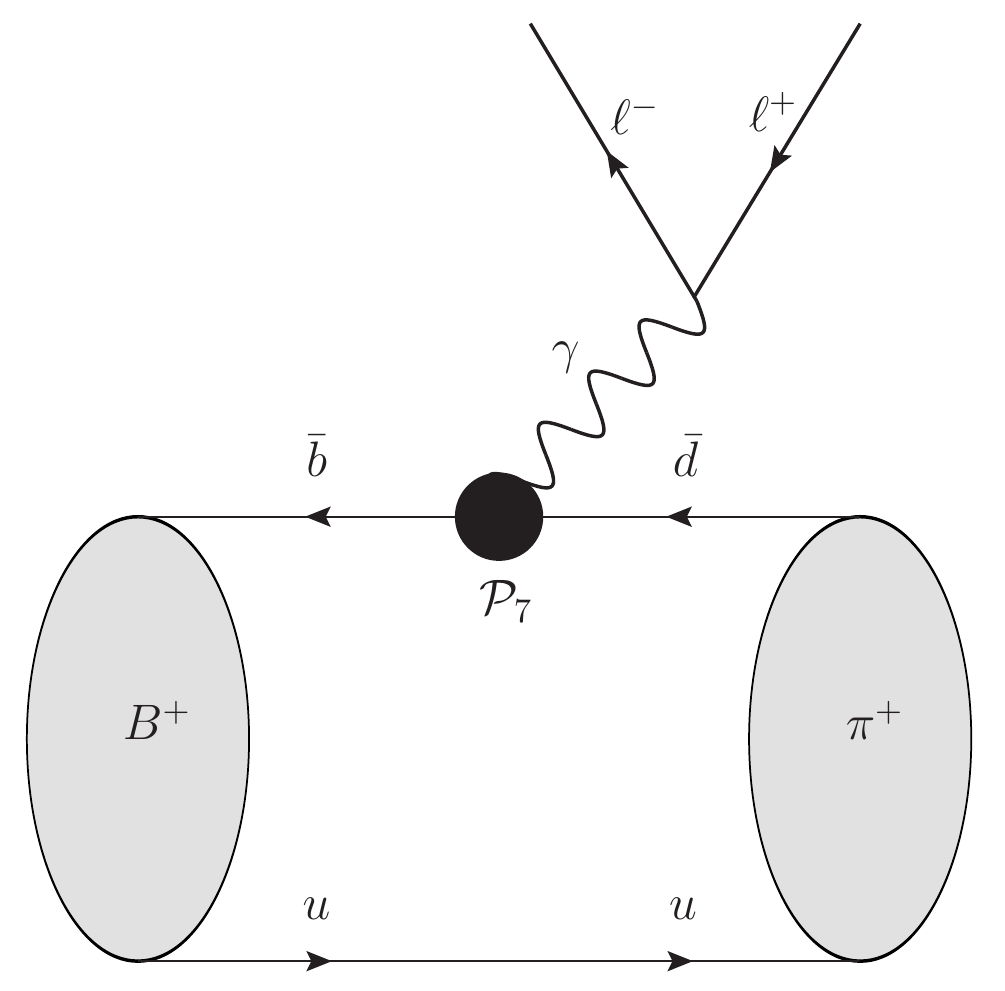}
\includegraphics[width=0.24\textwidth]{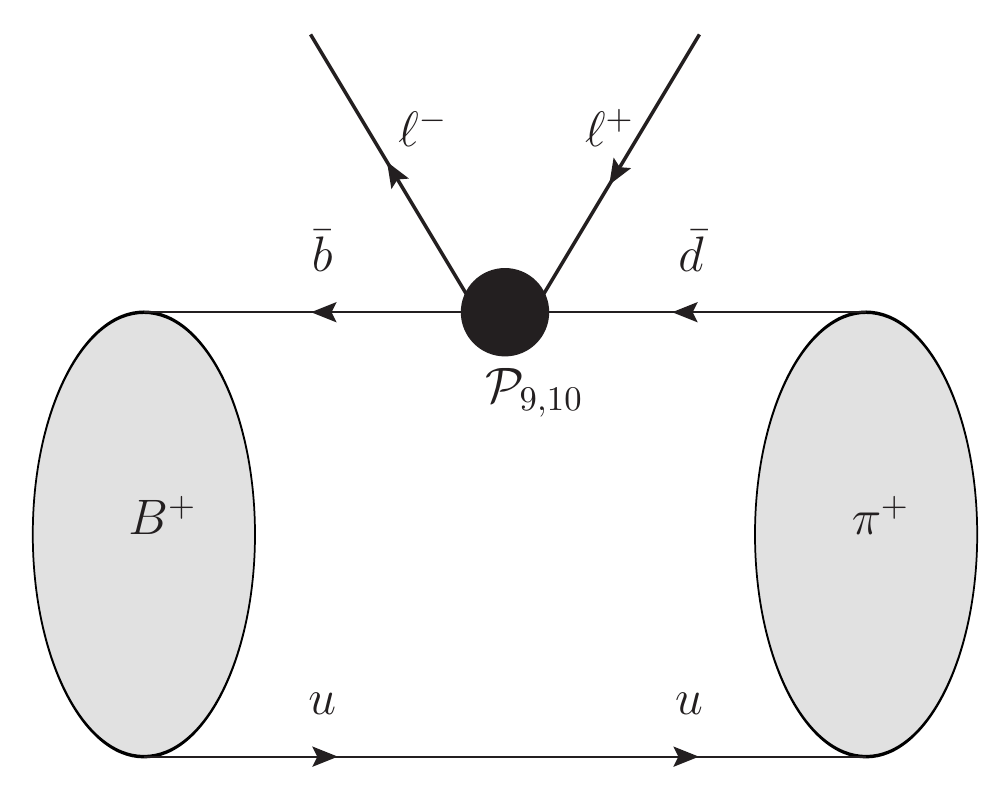}
}
\caption{Feynman diagrams of the $B^+ \to \pi^+ \ell^+ \ell^-$ decay.}
\label{fig:diag:1}
\end{figure}

Feynman diagrams for the $B^+ \to \pi^+ \ell^+ \ell^-$ decay are shown in Fig.~\ref{fig:diag:1}, 
where the left one denotes the ${\cal P}_{7\gamma}$ contribution, and the right one denotes 
the ${\cal P}_{9\ell}$ and ${\cal P}_{10\ell}$ contributions.
The matrix elements for the $B \to P$ transition, where $P$ is a pseudo-scalar meson, 
are expressed in terms of three transition form factors~\cite{Beneke:2000wa}: vector 
$f_+ (q^2)$, scalar $f_0 (q^2)$, and tensor $f_T (q^2)$, where $q^\mu = (p_B - k)^\mu$ 
is the four-momentum transferred to the lepton pair:
\begin{eqnarray}
&& \hspace{-11mm}
\langle P (k) | \bar p \gamma^\mu b | B (p_B) \rangle = f_+ (q^2)  
\label{eq:fp-f0-def} \\
&& \hspace{-7mm} \times 
\left [ p_B^\mu + k^\mu - \frac{m_B^2 - m_P^2}{q^2}\, q^\mu \right ]  
+ f_0 (q^2)\, \frac{m_B^2 - m_P^2}{q^2}\, q^\mu , 
\nonumber \\ 
% \end{eqnarray}
% 
% \begin{eqnarray}
&& \hspace{-11mm}
\langle  P (k) | \bar p \sigma^{\mu\nu} q_\nu b | B (p_B) \rangle =
 i\, \frac{f_T (q^2)}{m_B + m_P} 
\label{eq:fT-def} \\ 
&& \hspace{-7mm} \times 
\left [ \left ( p_B^\mu + k^\mu \right ) q^2 - q^\mu \left ( m_B^2 - m_P^2 \right) \right ] ,  
\nonumber 
\end{eqnarray}
where $m_B$ and $m_P$ are the $B$- and pseudo-scalar meson masses, respectively.

Taking into account the
sub-leading contributions, the  differential branching fraction 
is as follows~\cite{Ali:2020tjy}:

\begin{eqnarray}
&& \hspace*{-14mm}
\frac{d \mbox{Br} \left ( B \to P \ell^+ \ell^- \right )}{d q^2} = 
S_P \frac{2 G_F^2 \alpha_{\rm em}^2 \tau_B}{3 (4 \pi)^5 m_B^3}  
|V_{tb} V_{tp}^*|^2   
% \sqrt{\lambda(q^2) \left ( 1 - \frac{4 m_\ell^2}{q^2} \right )}  
\lambda^{3/2} (q^2) 
\nonumber \\ 
&& \times F^{BP} (q^2) \sqrt{1 - 4 m_\ell^2/q^2}, 
\label{eq:B-P-ell-ell-DBF} \\
&& \hspace*{-14mm} 
F^{BP} (q^2) = F_{97}^{BP} (q^2) + F_{10}^{BP} (q^2), 
\nonumber \\ 
&& \hspace*{-14mm}
F_{97}^{BP} (q^2) =   % \frac{2}{3} \, \lambda(q^2) 
\left ( 1 + \frac{2 m_\ell^2}{q^2} \right )
\Bigl | C_9^{\rm eff} (q^2) \, f_+^{BP} (q^2) 
\nonumber \\ 
&& 
+ \frac{2 m_b C_7^{\rm eff} (q^2)}{m_B + m_P} \, 
f_T^{BP} (q^2) + % \Bigr. 
% \nonumber \\ &&  
L_A^{BP} (q^2) + \Delta C^{BP}_V (q^2) \Bigr |^2, 
\nonumber \\
&& \hspace*{-14mm} 
F_{10}^{BP} (q^2) =   % \frac{2}{3} \, \lambda(q^2) 
\left ( 1 - \frac{4 m_\ell^2}{q^2} \right ) 
\left | C_{10}^{\rm eff} \, f_+^{BP} (q^2) \right |^2 
\nonumber \\ 
&&
+ \frac{6 m_\ell^2}{q^2} \, 
\frac{\left ( m_B^2 - m_P^2 \right)^2}{\lambda(q^2)} 
\left | C_{10}^{\rm eff} \, f_0^{BP} (q^2) \right |^2,   
\nonumber 
\end{eqnarray}
where $S_P$ is the isospin factor of the final meson 
($S_{\pi^\pm} = 1$ and $S_{\pi^0} = 1/2$ for the $\pi$-mesons, 
the case of our interest in this paper), 
% $p = s, d$ is flavor in $b \to p$ transition, 
$C_{7, 9, 10}^{\rm eff}$ are the effective Wilson coefficients 
including the NLO QCD corrections~\cite{Asatrian:2003vq}, 
$L_A^{BP} (q^2)$ is the Weak-Annihilation (WA) contribution, 
$\Delta C^{BP}_V (q^2)$ is the Long-Distance (LD) contribution, and 
\begin{equation}
\lambda (q^2) = \left ( m_B^2 + m_P^2 - q^2 \right )^2 - 4 m_B^2 m_P^2 ,
\label{eq:lambda-q2-def}
\end{equation}
is the kinematical function encountered in three-body decays (the triangle function).
Note that the differential branching fraction for the decay with the $\tau^+ \tau^-$-pair 
production differs from its counterparts with $e^+ e^-$ and $\mu^+ \mu^-$ due to 
the important role of the scalar form factor, $f_0 (q^2)$. 
In the electronic and muonic modes, its contribution is chirally suppressed by~$m_e^2$ 
and~$m_\mu^2$, respectively, while this no longer holds for the $\tau^+\tau^-$ case.

The WA contribution is calculated in the so-called Large Energy Effective Theory 
(LEET)~\cite{Beneke:2004dp} and has a significant effect for $q^2 \lesssim 1$~GeV$^2$ only,
so its inclusion makes sense for the $B^\pm \to \pi^\pm e^+ e^-$ and 
$B^\pm \to \pi^\pm \mu^+ \mu^-$ decays, but it is irrelevant for the 
$B^\pm \to \pi^\pm \tau^+ \tau^-$ case having the $q^2$-threshold above 12~GeV$^2$.

Two-particle decays, $B \to V \pi$,
%  where $V = \rho^0,\, \omega,\, \phi,\, J/\psi,\, \psi(2S)$ 
where~$V$ is a neutral vector meson, followed by the leptonic decay $V \to \ell^+ \ell^-$ 
determine the LD-contributions. They can be represented as follows~\cite{Hambrock:2015wka}:
\begin{eqnarray}
&& \hspace{-7mm}
\Delta C^{B\pi}_V = - 16 \pi^2\, 
% \nonumber \\
% && \times 
\frac{V_{ub}^{\phantom *} V_{ud}^* H^{(u)} + V_{cb}^{\phantom *} V_{cd}^* H^{(c)}}
     {V_{tb}^{\phantom *} V_{td}^*},
\label{eq:ldc1} \\ 
&& 
\hspace{-14mm}
H^{(p)}(q^2) = \sum_V 
\frac{\left ( q^2 - q_0^2 \right ) k_V f_V A^p_{BV\pi}}
     {(m_V^2 - q_0^2)(m_V^2 - q^2 - i m_V \Gamma^{\rm tot}_V)} 
% \nonumber \\
% && \hspace{-14mm} \times 
 ,
\label{eq:ldc2}
\end{eqnarray}
where $m_V$, $f_V$ and $\Gamma^{\rm tot}_V$ are the mass, decay constant and total decay 
width of the vector meson, respectively, $k_V$~is a valence quark content factor, 
$A^p_{BV\pi}$ ($p = u,\,c$) are the transition amplitudes, and the free parameter 
$q_0^2 = - 1.0$~GeV$^2$ is chosen to achieve a better convergence in the denominator 
of~(\ref{eq:ldc2}). The differential branching fraction~(\ref{eq:B-P-ell-ell-DBF}) 
involves three $B \to P$ form factors. They are scalar functions of~$q^2$,  
discussed for the $B \to \pi$ case in the next section.

\section{Form Factor Parametrizations}
\label{sec:FF-param}

Among the available parametrizations of the $B \to \pi$ transition form factors (FF) known 
in the literature, we chose those which are based on analyticity, crossing symmetry and 
the QCD dispersion relations. They are represented as a series in powers of the function 
$z (q^2, q_0^2)$ projecting~$q^2$ into the unit ellipse in the complex plane\footnote{
Parameter $q_0^2$ used here differs from the one in Eq.~(\ref{eq:ldc2}).}.

The first one is the Boyd-Grinstein-Lebed (BGL) parametrization~\cite{Boyd:1995sq} ($i = +,\, 0,\, T$):  
\begin{eqnarray}
&& \hspace{-1cm} 
f_i (q^2) = \frac{1}{P_i (q^2)\, \phi_i (q^2, q_0^2)} \sum\limits_{k = 0}^N
a^{(i)}_k \, z^k (q^2, q_0^2) ,  
\label{eq:BGL-param} \\
&& \hspace{-1cm} 
z (q^2, q_0^2) = \frac{\sqrt{m_+^2 - q^2} - \sqrt{m_+^2 - q_0^2}}
                      {\sqrt{m_+^2 - q^2} + \sqrt{m_+^2 - q_0^2}} , 
\label{eq:BGL-z-def} 
\end{eqnarray}
where $P_{i = +, T} (q^2) = z (q^2, m_{B^*}^2)$ and $P_0 (q^2) = 1$ are the Blaschke factors, 
$m_{B^*} = (5324.71 \pm 0.21)$~MeV is the vector $B^*$-meson mass~\cite{ParticleDataGroup:2022pth}, 
$\phi_i (q^2, q_0^2)$ is an outer function~\cite{Boyd:1995sq}, 
depending on three free parameters~$K_i$, $\alpha_i$, and~$\beta_i$, 
$m_+ = m_B + m_\pi$, and $q_0^2 = 0.65 \left (m_B - m_\pi \right )^2$. 
Expansion coefficients~$a^{(i)}_k$ are non-perturbative parameters,
which are determined either phenomenologically or by non-perturbative methods.

The second one is the Bourrely-Caprini-Lellouch (BCL) parametrization~\cite{Bourrely:2008za} ($i = +,\, T$):   
\begin{eqnarray}
&& \hspace{-1cm} 
f_i (q^2) = \frac{1}{1 - q^2/m_{B^*}^2}  
\nonumber \\ 
&& \hspace{-1cm} \times 
\sum\limits_{k = 0}^{N - 1} b^{(i)}_k 
\Biggl [ z^k (q^2, q_0^2) - (-1)^{k - N} \frac{k}{N}\, z^N (q^2, q_0^2) \Biggr ],  
\label{eq:BCL-f-pT} \\
&& \hspace{-1cm} 
f_0 (q^2) = \sum\limits_{k = 0}^{N - 1} b^{(0)}_k z^k (q^2, q_0^2), 
\label{eq:BCL-f-0} \\
&& \hspace{-1cm} 
q_0^2 = m_+ ( \sqrt{m_B} - \sqrt{m_\pi} )^2 .
\label{eq:BCL-q0-def}
\end{eqnarray}
Here, $z (q^2, q_0^2)$ is the same as in~Eq.~(\ref{eq:BGL-z-def}) 
and the form factors are calculated by truncating the series at $N~=~4$. 

The third type is the modified Bourrely-Caprini-Lellouch (mBCL) 
parametrization~\cite{Leljak:2021vte} ($i = +,\, T$):   
\begin{eqnarray}
&& \hspace{-1cm} 
f_i (q^2) = \frac{f_i (q^2 = 0)}{1 - q^2/m_{B^*}^2} 
%\nonumber \\ 
\label{eq:mBCL-param} \\
&& \hspace{-1cm} \times 
\Biggl [ 1 + \sum\limits_{k = 1}^{N - 1} b^{(i)}_k 
\Bigl ( \bar z_k ( q^2, q_0^2 ) - (-1)^{k - N} \frac{k}{N} \, \bar z_N ( q^2, q_0^2 ) \Bigr ) 
\Biggr] ,
\nonumber \\
%\label{eq:mBCL-param} \\
&& \hspace{-1cm} 
f_0 (q^2) = \frac{f_+ (q^2 = 0)}{1 - q^2/m_{B_0}^2} 
\Biggl [ 1 + \sum\limits_{k = 1}^N b^{(0)}_k \bar z_k ( q^2, q_0^2 ) \Biggr ],
\end{eqnarray}
where $\bar z_k (q^2, q_0^2) = z^k (q^2, q^2_0) - z^k (0, q_0^2)$. 
The function $z (q^2, q_0^2)$ is defined in~Eq.~(\ref{eq:BGL-z-def}) 
and $q_0^2$ takes the optimal value~(\ref{eq:BCL-q0-def}).                             
Here, unlike other types of the $f_0 (q^2)$ parametrizations, 
this form factor has a pole but at higher~$q^2$~--- 
at the scalar $B_0$-meson mass squared, $m_{B_0}^2$. 
This state is not yet observed experimentally and its mass 
is taken from theory. We set $m_{B_0}= 5.54$~GeV, as was used 
in the determination of the expansion coefficients~$b^{(0)}_k$~\cite{Leljak:2021vte}.   

Note that the Dispersion Matrix (DM) method was suggested in~\cite{DiCarlo:2021dzg}  % [69] 
to describe the FFs by using also analyticity, crossing symmetry and the QCD dispersion relations. 
This method is based on the non-perturbative determination of the dispersive bounds and describes 
in a model-independent way the FFs in the full kinematical range, starting from existing Lattice 
QCD data at large momentum transfer, without a series expansion in powers of $z (q^2, q_0^2)$. 
It was already applied to the semileptonic $B \to \pi \ell \nu_\ell$ decays~\cite{Martinelli:2022tte}  % [70] 
and can be also used for the analysis of semileptonic FCNC $B$-meson decays.

\section{Numerical Analysis of the $B^+ \to \pi^+ \tau^+ \tau^-$ Decay}
\label{sec:numerics}

\subsection{Perturbative Contribution}
\label{ssec:pert-contribution}

\begin{figure}[htb]
\begin{center}
\includegraphics[width=0.35\textwidth]{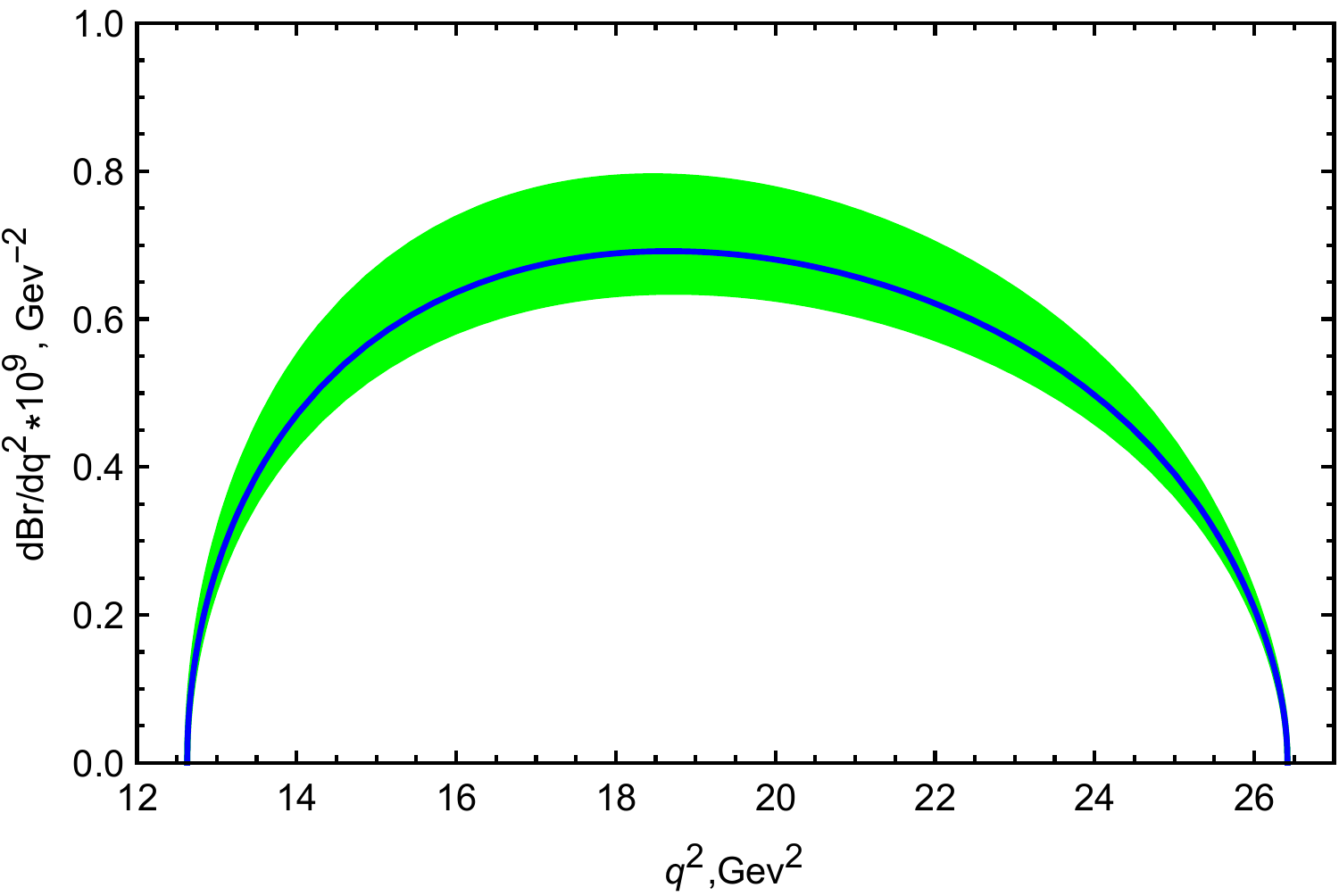} \\
\includegraphics[width=0.35\textwidth]{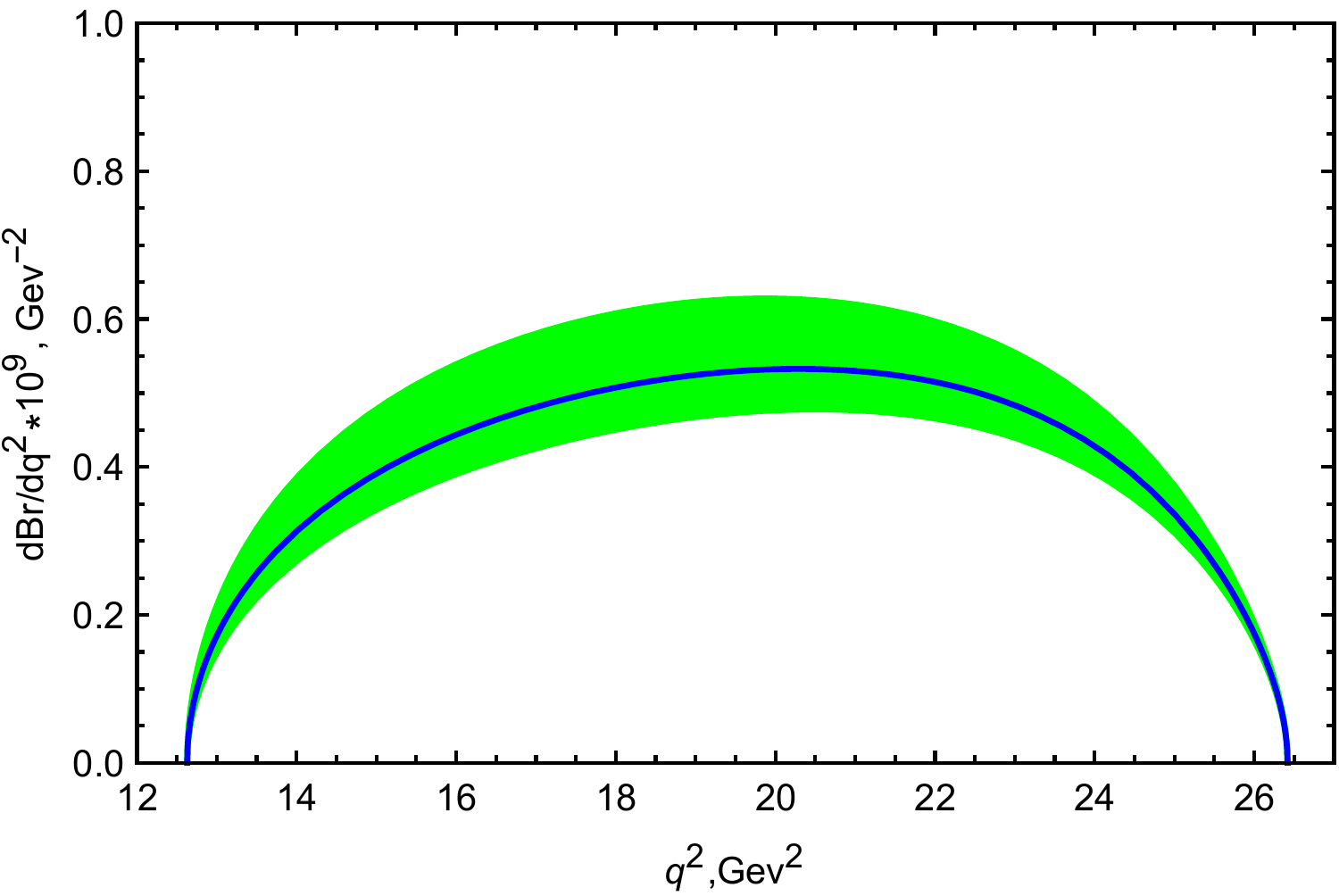} \\
\includegraphics[width=0.35\textwidth]{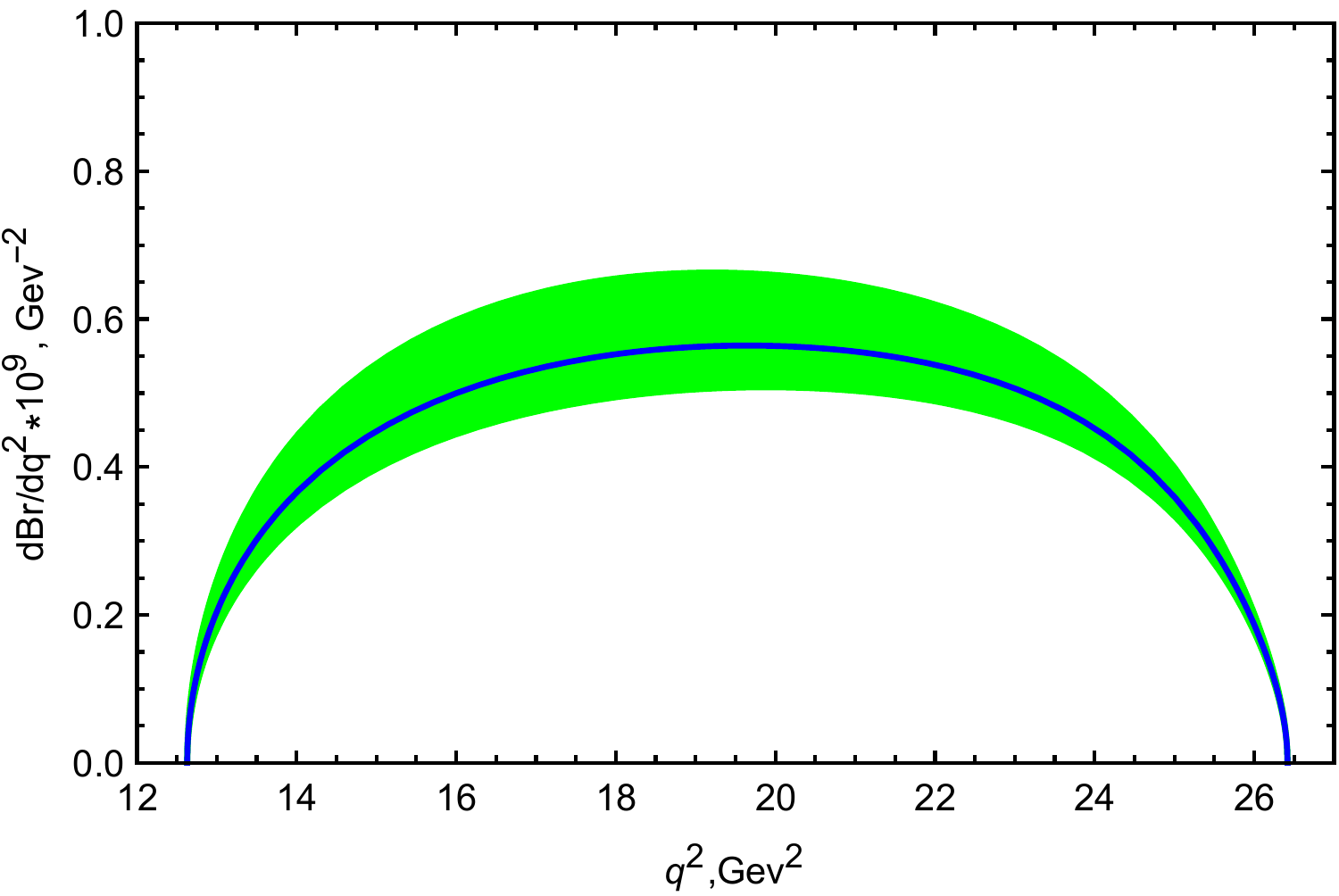}
\end{center}
\caption{\label{fig:lr}
The dilepton invariant-mass distribution for $B^+ \to \pi^+ \tau^+ \tau^-$ decay 
for the BGL (top), BCL (center) and mBCL (bottom) parametrizations of the form factors. 
The green areas indicate the uncertainty due to the factorization scale, FF expansion 
coefficients and CKM matrix element~$V_{td}$.
}
\end{figure}

The distribution in the tau-pair invariant mass calculated in perturbation theory 
for three types of form factor parametrizations is presented in Fig.~\ref{fig:lr}.
The spread shown in these distributions reflect the convoluted uncertainties 
in the scale parameter~$\mu$, entering via the Wilson coefficients by varying it in the 
range $m_b/2 \leq \mu \leq 2 m_b$, and the input value of the CKM matrix element 
$V_{td} = (8.54 \pm 0.30) \times 10^{-3}$~\cite{ParticleDataGroup:2022pth}.
Numerical results for the total branching fraction for the three FF parametrizations used 
in this work are consistent with each other within uncertainties as shown in Table~\ref{tab:Br}. 
In working out the numerical values, the expansion coefficients of the BGL parametrization 
are taken from~\cite{Ali:2013zfa}, where the data on the CC-process  $B \to \pi \ell \nu_\ell$ 
decay are fitted, and the relations between the $B \to K$ and $B \to \pi$ form factors are used. 
The values of the BCL parametrization coefficients were obtained within the framework 
of Lattice QCD (LQCD)~\cite{FermilabLattice:2015cdh,FermilabLattice:2015mwy}.
The values of the mBCL parametrization coefficients are
obtained by the combined use of the Light-Cone Sum Rules (LCSR) and LQCD~\cite{Leljak:2021vte}. 
All the expansion coefficients are collected in~\ref{app:1}. The entries in Table~\ref{tab:Br} 
are also consistent with the existing theoretical predictions in the literature, for example, 
${\rm Br}^{\rm FG}_{\rm th} (B^+ \to \pi^+ \tau^+ \tau^-) = (7.0 \pm 0.7) \times 10^{-9}$~\cite{Faustov:2014zva} 
and 
${\rm Br}^{\rm WX}_{\rm th} (B^+ \to \pi^+ \tau^+ \tau^-) = (6.0^{+2.6}_{-2.1}) \times 10^{-9}$~\cite{Wang:2012ab}.

\begin{table}[tb]
\caption{
Theoretical predictions for the $B^+ \to \pi^+ \tau^+ \tau^-$ total branching fraction, obtained for the three indicated FF parametrizations.
}
\label{tab:Br}
\centerline{
\begin{tabular}{|c|c|c|c|}
\hline
& BGL & BCL & mBCL \\
\hline
${\rm Br}_{\rm th} \times 10^9$ & 
$7.56^{+0.74}_{-0.43}$ & $6.00^{+0.81}_{-0.49}$ & $6.28^{+0.76}_{-0.46}$ \\
\hline
\end{tabular}
}
\end{table}

\begin{figure}[htb]
\begin{center}
\includegraphics[width=0.35\textwidth]{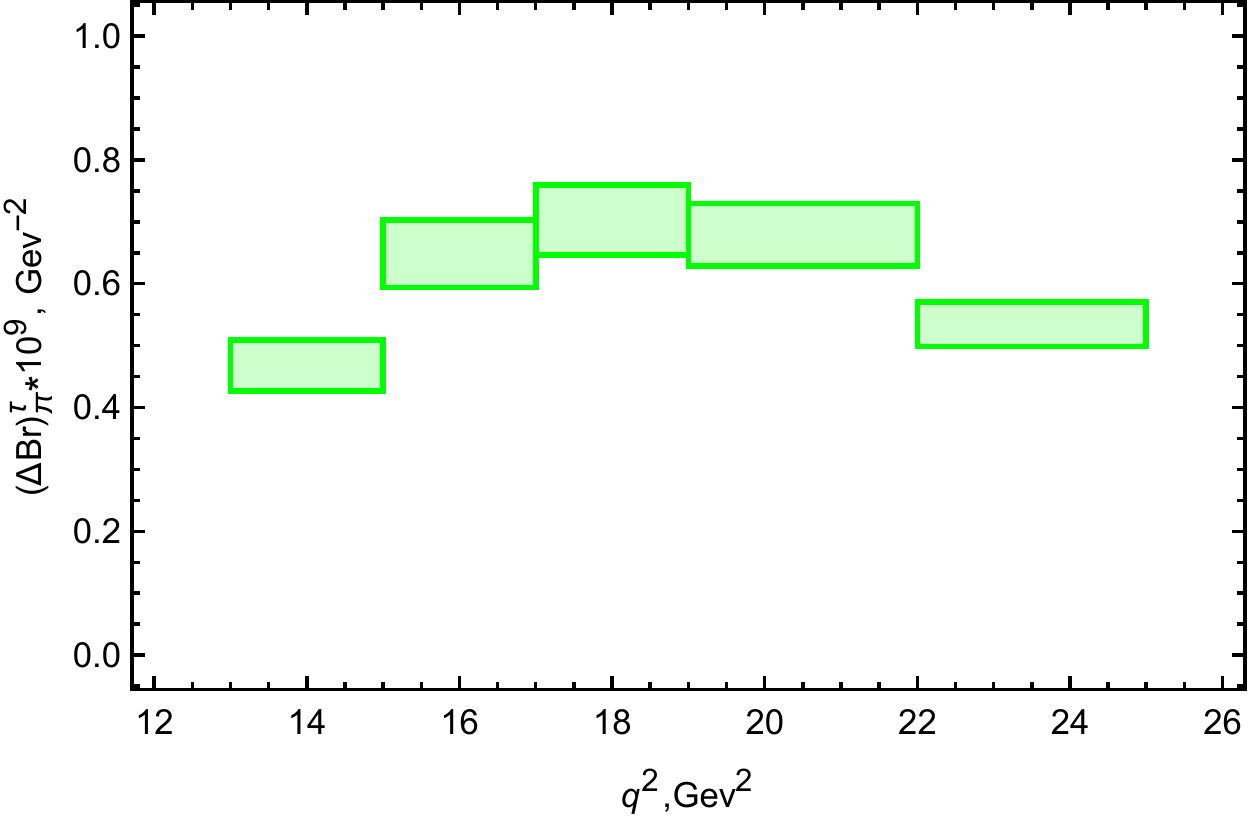} \\
\includegraphics[width=0.35\textwidth]{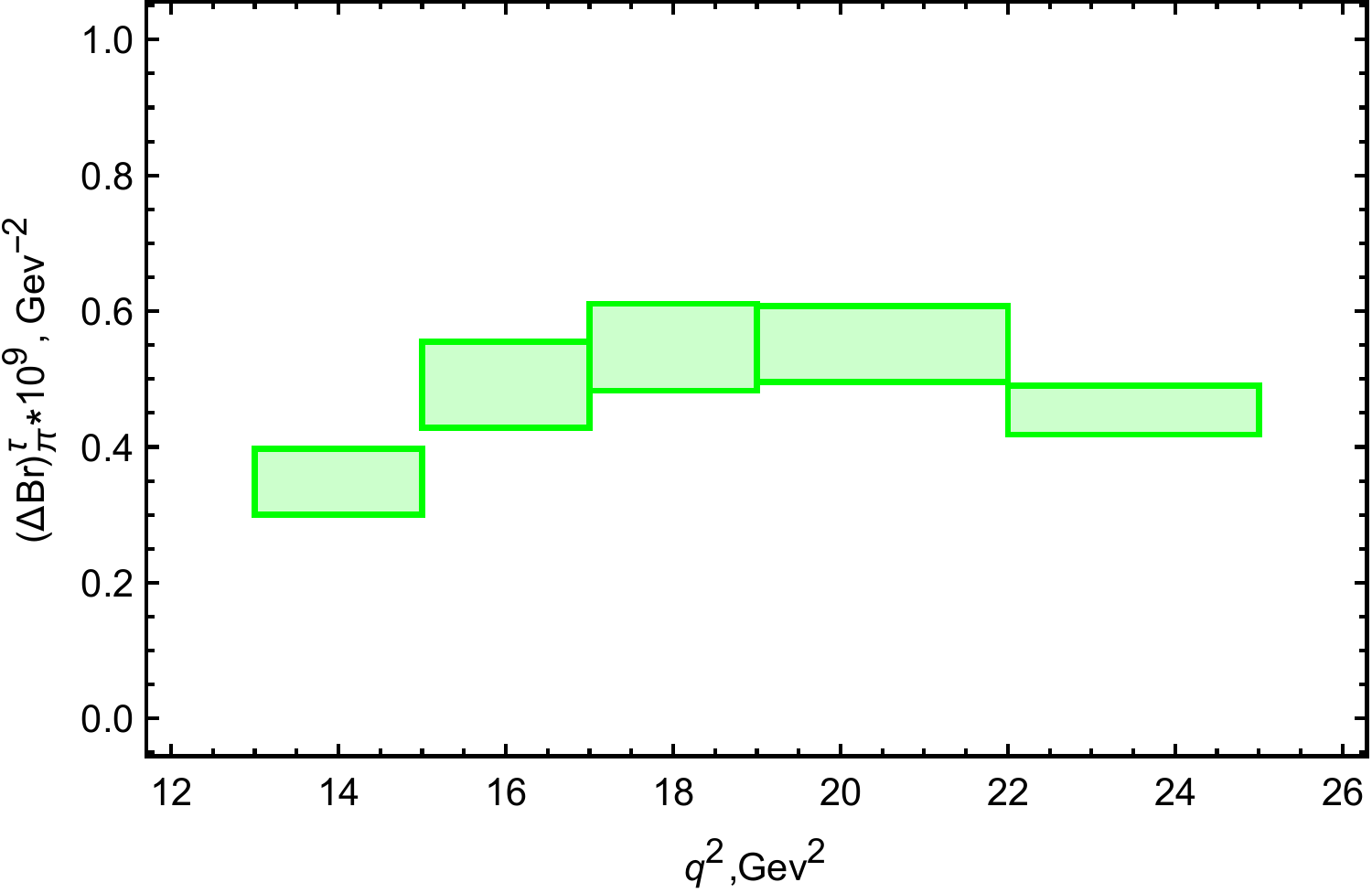} \\
\includegraphics[width=0.35\textwidth]{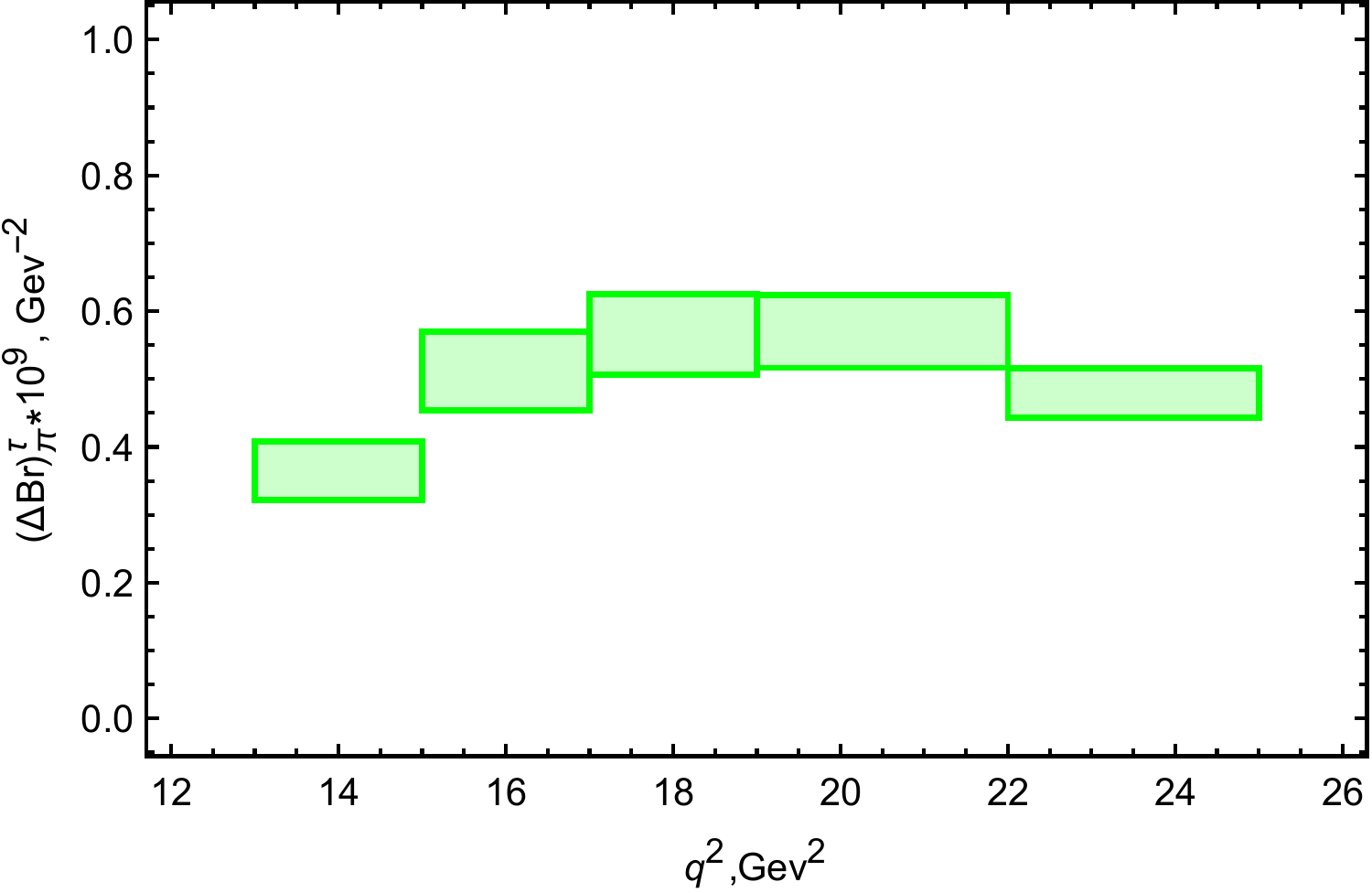}

\caption{\label{fig:bin}
Partial branching fraction of the $B^+ \to \pi^+ \tau^+ \tau^-$ decay, 
$(\Delta {\rm Br})_\pi^\tau (q_{\rm min}^2, q_{\rm max}^2)$, in bins of ditauon invariant 
mass squared for the BGL (top), BCL (center) and mBCL (bottom) form factor parametrizations. 
}
\end{center}
\end{figure}

It is customary to compare data and theoretical distributions in bins of $q^2$: 
\begin{equation}
(\Delta {\rm Br})_\pi^\tau (q_1^2, q_2^2) \equiv
\int_{q_1^2}^{q_2^2} dq^2\, \frac{d{\rm Br} (B^+ \to \pi^+ \tau^+ \tau^-)}{dq^2} . 
\label{eq:DeltaBr-def} 
\end{equation}
To that end, we plot the theoretical results for the partial branching ratio
$(\Delta {\rm Br})_\pi^\tau (q_1^2, q_2^2) $ in bins of the ditauon invariant-mass squared 
using the three FF parametrization in Fig.~\ref{fig:bin}, and collect the corresponding values 
of the partial branching fractions, integrated over the indicated ranges, in Table~\ref{tab:tau}. 
% We take into account 
%the uncertainties due to the factorization scale variation, CKM matrix element~$V_{td}$ and the uncertainties in the parameters intrinsic to the form factors.
For comparison, the Lattice results~\cite{FermilabLattice:2015mwy} are also shown in the last column.
The errors shown are from the CKM matrix element, form factors, variation of the high and low matching 
scales, and the quadrature sum of all other contributions, respectively. 
We note that the BGL parametrization predictions are in good agreement with the Lattice-based estimates, 
both of which are, however, systematically higher than the predictions based on the BCL and mBCL ones 
in each bin. The same also holds for the total branching fraction (see Table~\ref{tab:Br}). 

\begin{table}[tb]
{\scriptsize
\begin{center}
\caption{\label{tab:tau}
Partial branching ratios for the $B^+ \to \pi^+ \tau^+ \tau^-$ decay,
$(\Delta {\rm Br})_\pi^\tau (q_{\rm min}^2, q_{\rm max}^2)$, obtained using 
the BGL, BCL and mBCL FF parametrizations in comparison with the Lattice QCD 
predictions~\cite{FermilabLattice:2015mwy}. Invariant mass squared,~$q^2$, 
is given in units of GeV$^2$. Errors in the last column obtained 
by the Lattice QCD calculations are explained in the text.  
}
% \hspace{-0.7cm}
\begin{tabular}{|c|c|c|c|c|}
     \hline
& \multicolumn{4}{c|}{$10^9 \times (\Delta {\rm Br})_\pi^\tau$}  \\
\hline
[$q_{\rm min}^2, q_{\rm max}^2$] & BGL & BCL &  mBCL &  Lattice-QCD\cite{FermilabLattice:2015mwy}  \\
\hline
[13.0, 15.0] & $0.91^{+0.11}_{-0.06}$ & $0.67^{+0.12}_{-0.07}$ & $0.71^{+0.11}_{-0.06}$ & -  \\[0.5mm]
\hline
[15.0, 17.0] & $1.27^{+0.14}_{-0.08}$ & $0.95^{+0.16}_{-0.09}$ & $1.00^{+0.15}_{-0.09}$ & $1.11 ( 7,  8, 2, 4)$ \\[0.5mm]
\hline
[17.0, 19.0] & $1.37^{+0.14}_{-0.08}$ & $1.06^{+0.16}_{-0.10}$ & $1.10^{+0.15}_{-0.09}$ & $1.25 ( 8,  8, 2, 3)$ \\[0.5mm]
\hline
[19.0, 22.0] & $2.00^{+0.19}_{-0.11}$ & $1.62^{+0.21}_{-0.13}$ & $1.67^{+0.20}_{-0.12}$ & $1.93 (12, 10, 4, 5)$ \\[0.5mm]
\hline
[22.0, 25.0] & $1.58^{+0.13}_{-0.09}$ & $1.34^{+0.13}_{-0.09}$ & $1.42^{+0.13}_{-0.09}$ & $1.59 (10,  7, 4, 4)$ \\[0.5mm]
\hline
\end{tabular}
\end{center}
}
\end{table}

\subsection{Long-Distance Contributions}
\label{ssec:LD-contribution}

Since the $q^2$-threshold in the $B^+ \to \pi^+ \tau^+ \tau^-$ decay is 
$4 m_\tau^2 = 12.6$~GeV$^2$, the tauonic invariant-mass distribution would include 
the $\psi (2S)$-meson and higher charmonia decaying into the $\tau^+ \tau^-$-pair,
estimated below. 

For the $B^+ \to \pi^+ \tau^+ \tau^-$ decay, contribution from the $\psi (2S)$-meson, 
the total branching fractions for the $B^+ \to \pi^+ \psi (2S)$ and 
$\psi (2S) \to \tau^+ \tau^-$ decays are as follows~\cite{ParticleDataGroup:2022pth}:
\begin{eqnarray}
&& \hspace{-11mm} 
{\rm Br} (B^+ \to \pi^+ \psi (2S)) = \left ( 2.44 \pm 0.30 \right ) \times 10^{-5},  
\label{eq:Br-Bp-to-pip-psi2S} \\
&& \hspace{-11mm} 
{\rm Br} (\psi (2S) \to \tau^+ \tau^-) = \left ( 3.1 \pm 0.4 \right ) \times 10^{-3} , 
\label{eq:Br-psi2S-tau-tau} 
\end{eqnarray}
which yield the following product branching ratio:  
% of $B^+ \to \pi^+ \psi(2S) \to \pi^+ \tau^+ \tau^-$:
% 
\begin{equation}
{\rm Br} (B^+ \to \pi^+ \psi(2S) \to \pi^+ \tau^+ \tau^-) = \left ( 7.6 \pm 1.3 \right ) \times 10^{-8} . 
\label{eq:Br-Bp-pip-psi2S} 
\end{equation}
Being of order $10^{-7}$, the $\psi(2S)$-contribution strongly modifies the SD-based 
ditauon-mass spectrum but, as $\psi(2S)$-meson is a narrow resonance with 
$M_{\psi(2S)}^2 \simeq 13.6$~GeV$^2$ and has the decay width 
$\Gamma_{\psi(2S)} = \left ( 294 \pm 8 \right )$~keV~\cite{ParticleDataGroup:2022pth}, it affects only 
the $q^2$-region in the vicinity of the $B^+ \to \pi^+ \tau^+ \tau^-$ threshold. 
Experimentally, this contribution can be largely reduced by putting kinematical cuts, say, 
$q^2 \ge 15$~GeV$^2$.  
 
The next vector $c\bar{c}$ resonance is the $\psi (3S)$-meson, also known as $\psi (3770)$. 
The total branching fractions of the $B^+ \to \pi^+ \psi (3S)$ and $\psi (3S) \to \tau^+ \tau^-$ 
decays are not yet known experimentally. We can estimate the pionic $B$-meson decay ratio  
by using the kaonic $B$-meson decay modes, $B^+ \to K^+ \psi (2S)$ and $B^+ \to K^+ \psi (3S)$,  
which have been measured:   
${\rm Br} (B^+ \to K^+ \psi (2S)) = \left ( 6.24 \pm 0.20 \right ) \times 10^{-4}$ and 
${\rm Br} (B^+ \to K^+ \psi (3S)) = \left ( 4.3 \pm 1.1 \right ) \times 10^{-4}$~\cite{ParticleDataGroup:2022pth}.  
The branching fraction for the decay $B^+ \to \pi^+ \psi (3S)$ can be found with the help 
of the (approximate) $SU (3)_F$ relation:
\begin{equation}
\frac{{\rm Br} (B^+ \to \pi^+ \psi (3S))}{{\rm Br} (B^+ \to K^+ \psi (3S))} \simeq 
\frac{{\rm Br} (B^+ \to \pi^+ \psi (2S))}{{\rm Br} (B^+ \to K^+ \psi (2S))} ,  
\label{eq:psi2S-psi3S-prod-relation}
\end{equation}
where ${\rm Br} (B^+ \to \pi^+ \psi (2S))$ is presented in~(\ref{eq:Br-Bp-to-pip-psi2S}). 
Taking~(\ref{eq:psi2S-psi3S-prod-relation}) as a good approximation, we get: 
\begin{equation}
{\rm Br} (B^+ \to \pi^+ \psi(3S)) = \left ( 1.7 \pm 0.5 \right ) \times 10^{-5} . 
\label{eq:Br-Bp-pip-psi3S} 
\end{equation}
However, in contrast to the  narrow $\psi (2S)$-meson, $\psi (3770)$ is a broad resonance which decays 
mainly to $D^+ D^-$ and $D^0 \bar D^0$, so its purely leptonic decay modes are strongly 
suppressed. To see this suppression numerically for $\psi (3S) \to \tau^+ \tau^-$, we use the 
lepton flavor universality, obeyed by QED and QCD, and the experimentally measured branching fraction 
${\rm Br} (\psi (3S) \to e^+ e^-) = \left ( 9.6 \pm 0.7 \right ) \times 10^{-6}$~\cite{ParticleDataGroup:2022pth}. 
The branching ratios ${\rm Br} (\psi(3S) \to e^+ e^-)$ and  ${\rm Br} (\psi(3S) \to \tau^+ \tau^-)$ differ 
from each other only by the phase space factor and hence their relative rates follow the kinematic relation:
\begin{equation}
\frac{{\rm Br} (\psi(3S) \to \tau^+ \tau^-)}{{\rm Br} (\psi(3S) \to e^+ e^-)} = 
\frac{\lambda ( M_{\psi(3S)}, m_\tau, m_\tau )}{\lambda ( M_{\psi(3S)}, m_e, m_e )},
\label{eq:psi2S-psi3S-decay-relation}
\end{equation}
where $\lambda (M, m, m) = M \sqrt{M^2 - 4 m^2}$~\cite{Bycking:1973}.
Taking the masses into account: $M_{\psi (3S)} = \left ( 3773.7 \pm 0.4 \right )$~MeV, 
$m_e = 0.511$~MeV, and $m_\tau = \left ( 1776.86 \pm 0.12 \right )$~MeV~\cite{ParticleDataGroup:2022pth}, 
we obtain:
\begin{eqnarray}
{\rm Br} (\psi(3S) \to \tau^+ \tau^-) = \left ( 3.2 \pm 0.2 \right ) \times 10^{-6} , 
\label{eq:Br-psi3S-tau-tau} 
\end{eqnarray}
which in turn yields the LD branching fraction ${\rm Br} (B^+ \to \pi^+ \tau^+ \tau^-)$ 
from the $\psi(3S)$ resonance: 
\begin{equation}
{\rm Br} (B^+ \to \pi^+ \psi(3S) \to \pi^+ \tau^+ \tau^-) = \left ( 5.4 \pm 1.9 \right ) \times 10^{-11} . 
\label{eq:Br-B-psi3S-tau-tau} 
\end{equation}
This value is three orders of magnitude smaller than the similar decay rate of the $\psi(2S)$-meson~(\ref{eq:Br-Bp-pip-psi2S}). 
Comparing with the SD (perturbative) contribution (see Table~\ref{tab:tau}), which is of order of $10^{-9}$, 
the $\psi (3770)$ contribution, ${\rm Br} (B^+ \to \pi^+ \psi(3S) \to \pi^+ \tau^+ \tau^-)$, 
is subdominant, comparable to the current perturbative errors. 

There are yet more vector charmonium resonances with masses above the $\psi (3S)$-meson mass, 
$\psi (4040)$, $\psi (4160)$, $\psi (4230)$, $\psi (4360)$, and $\psi (4415)$, which also 
have purely leptonic decay modes. However, as they decay strongly into the $D \bar D$-pair etc., 
their electronic or muonic decay rates are also of order of $10^{-5}$~\cite{ParticleDataGroup:2022pth}, 
similar to the case of the $\psi (3S)$-meson, as shown in Table~\ref{tab:vector-charmonia}. 
It follows that their contributions to the $B^+ \to \pi^+ \tau^+ \tau^-$ branching fraction 
are of the same order of magnitude as from the~$\psi (3S)$-meson. Consequently, we drop 
the contribution from all the strongly decaying charmonium resonances ($ \psi(3S)$ and higher), 
and consider the LD-contribution from the narrow $\psi (2S)$-meson only.

\begin{table*} 
\caption{
Experimental data~\cite{ParticleDataGroup:2022pth} on vector charmonia 
with masses above the open charm threshold. The branching fraction of 
the $\psi (4360) \to e^+ e^-$ decay follows from the electronic decay 
width $\Gamma_{ee} = \left ( 11.6^{+5.0}_{-4.4} \pm 1.9 \right )$~eV in which 
the errors are added in quadrature. In getting the $V \to \tau^+ \tau^-$ 
branching fractions, Eq.~(\ref{eq:psi2S-psi3S-decay-relation}) is used. 
} 
\label{tab:vector-charmonia} 
\begin{center}
\begin{tabular}{|c|c|c|c|c|c|} 
\hline 
    $V$       &  $M_V$\, [MeV]   & $\Gamma_V$\, [MeV] & $10^5 \times {\rm Br} (B^+ \to V K^+)$ & $10^6 \times {\rm Br} (V \to e^+ e^-)$ & $10^6 \times {\rm Br} (V \to \tau^+ \tau^-)$ \\ \hline 
$\psi (4040)$ & $4039   \pm 1$   &    $80 \pm 10$     &          $1.1 \pm 0.5$                 &         $10.7 \pm  1.6$                &         $ 5.1 \pm   0.8$                     \\ \hline 
$\psi (4160)$ & $4191   \pm 5$   &    $70 \pm 10$     &          $ 51 \pm  27$                 &         $ 6.9 \pm  3.3$                &         $ 3.7 \pm   1.7$                     \\ \hline 
$\psi (4230)$ & $4222.7 \pm 2.6$ &    $49 \pm  8$     &                                        &         $  31 \pm   28$                &         $  17 \pm    15$                     \\ \hline 
$\psi (4360)$ & $4372   \pm 9$   &   $115 \pm 13$     &                                        &         $0.10 \pm 0.05$                &         $0.06 \pm  0.03$                     \\ \hline 
$\psi (4415)$ & $4421   \pm 4$   &    $62 \pm 20$     &          $2.0 \pm 0.8$                 &         $ 9.4 \pm  3.2$                &         $ 5.6 \pm   1.9$                     \\ \hline 
\end{tabular} 
\end{center}
\end{table*}

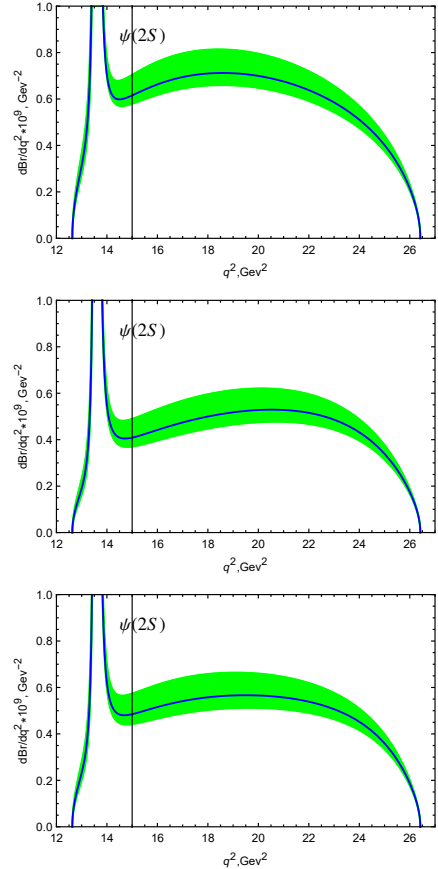
\begin{figure}[htb]
\begin{center}
\begin{picture}(100,100)(70,0)
\put(30, 0){\includegraphics[width=0.35\textwidth]{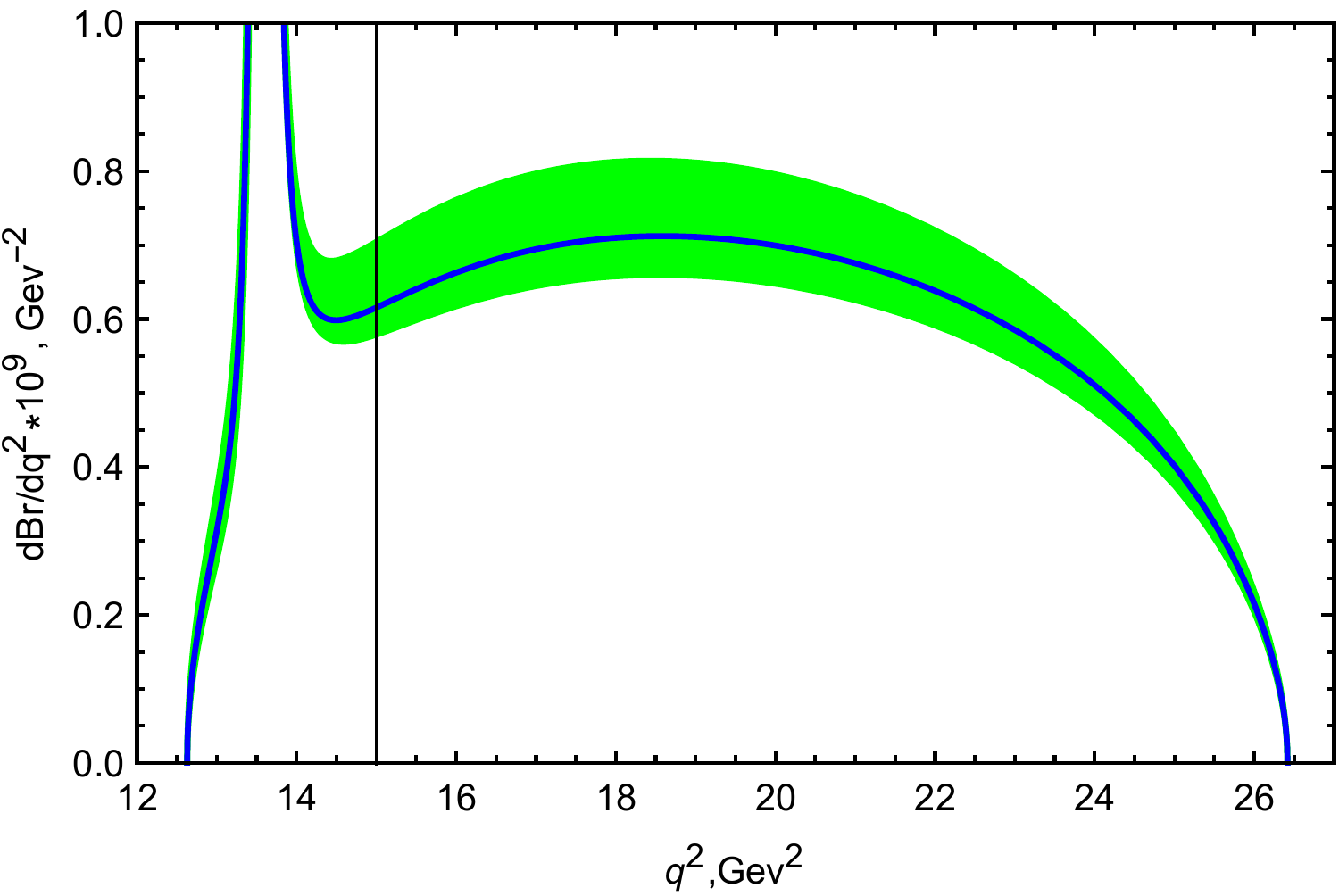}}
\put(70,90){{\scriptsize $\psi (2S)$}}
\end{picture}
\end{center} 
%
% \quad
% 
\begin{center}
\begin{picture}(100,100)(70,0)
\put(30, 0){\includegraphics[width=0.35\textwidth]{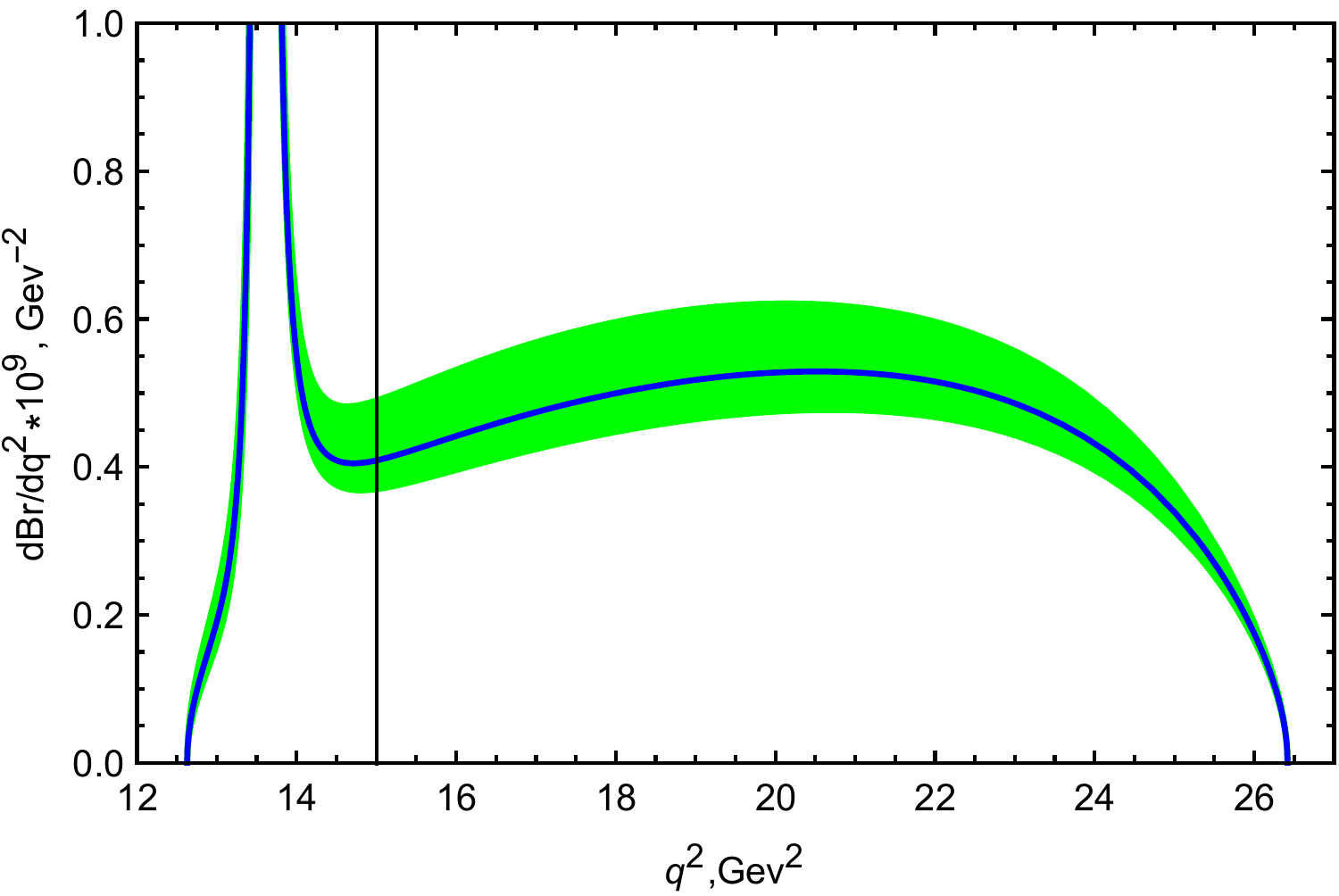}}
\put(70,90){{\scriptsize $\psi (2S)$}}
\end{picture}
\end{center} 
\begin{center}
\begin{picture}(100,100)(70,0)
\put(30, 0){\includegraphics[width=0.35\textwidth]{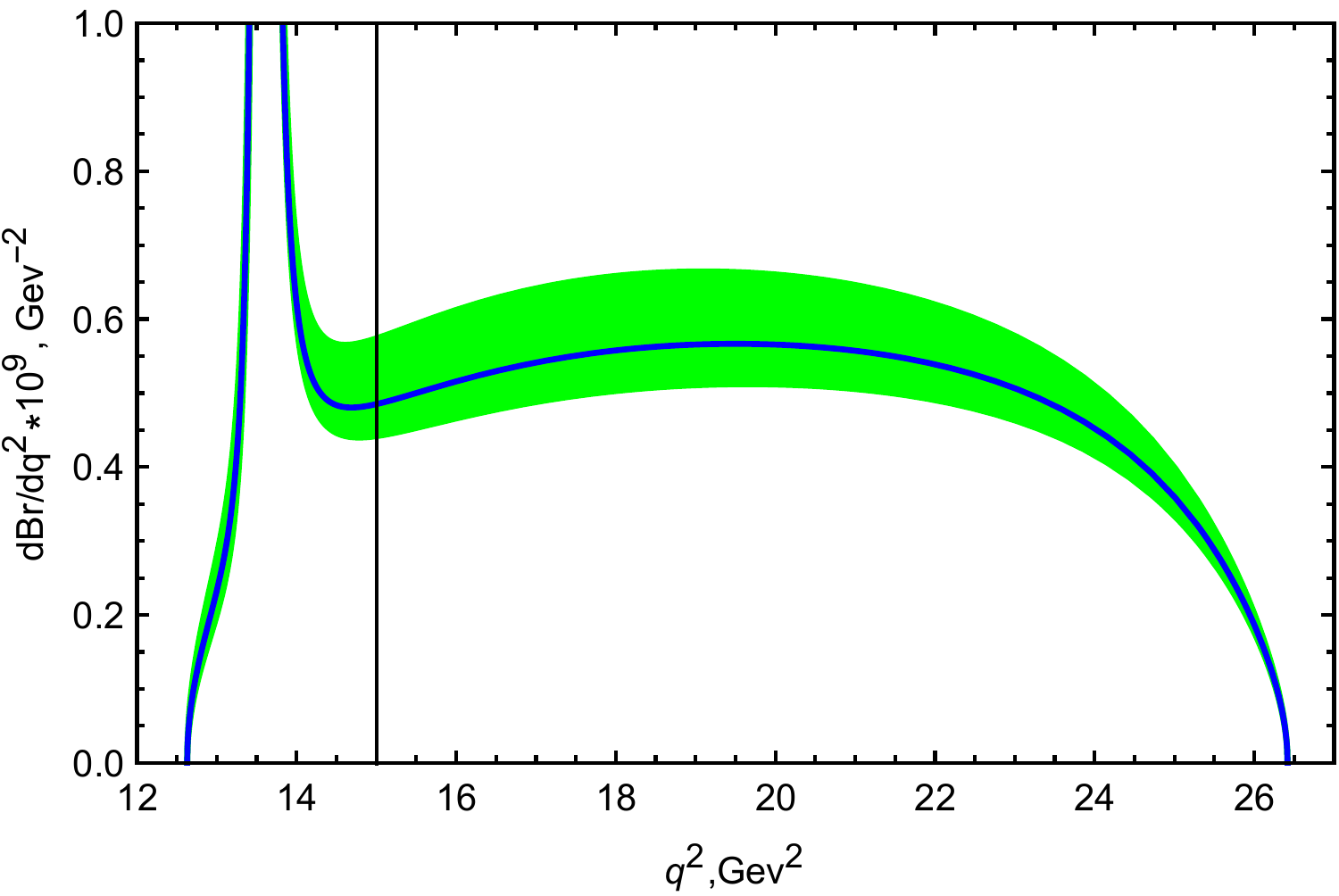}}
\put(70,90){{\scriptsize $\psi (2S)$}}
\end{picture}
\end{center} 
\caption{
The ditauon invariant mass distribution with the long-distance contribution from 
the $\psi (2S)$-meson for the BGL (upper plot), BCL (central plot) and mBCL (lower plot) 
parametrizations. The green areas indicate the uncertainty due to the factorization 
scale, FF expansion coefficients and the CKM matrix element~$V_{td}$. The vertical solid 
line at $q^2 = 15$~GeV$^2$ indicates the kinematical cut to exclude the $\psi (2S)$ contribution. 
}
\label{fig:tauwres}
\end{figure}

Since the LD contributions~(\ref{eq:ldc2}) depend on the choice of the amplitude 
phases~$\delta^{(u)}_{\psi (2S)}$ and~$\delta^{(c)}_{\psi (2S)}$, we present the  
total branching fraction of the $B^+ \to \pi^+ \tau^+ \tau^-$ decay, including 
the $\psi(2S)$ LD-contribution, and its dependence on the assumed values of the strong phases 
in Tables~\ref{tab:taudeltaCBGL}, \ref{tab:taudeltaCBCL} and~\ref{tab:taudeltaCmBCL} 
for the BGL, BCL and mBCL parametrizations of the form factors, respectively. 
As can be seen, the variation of the branching fraction on the strong phases 
is not very marked, and is similar to the errors shown from the SD contribution. 
The central value including the LD contribution is given for  
$\delta^{(u)}_{\psi (2S)} = 0$ and $\delta^{(c)}_{\psi (2S)} = 3\pi/4$.
The ditauon invariant mass distribution including the LD contribution from the $\psi(2S)$-meson 
is presented in Fig.~\ref{fig:tauwres} for the BGL, BCL and mBCL form factors.
The vertical solid line at $q^2 = 15$~GeV$^2$ in the plots indicates the kinematical cut 
to exclude the dominant $\psi (2S)$ contribution. 
\begin{table}[tb]
\begin{center}
\caption{\label{tab:taudeltaCBGL}
The total branching fraction for the $B^+ \to \pi^+ \tau^+ \tau^-$ decay in the BGL parametrization 
including the LD contribution from the $\psi(2S)$-meson for the various assumed values 
of the amplitude phases. SDC means the short-distance (perturbative) contribution.
}
\begin{tabular}{|c|c|c|}
\hline
$\delta^{(u)}_{\psi (2S)}$ & $\delta^{(c)}_{\psi (2S)}$ & ${\rm Br} (B^+ \to \pi^+ \tau^+ \tau^-) \times 10^{-9}$ \\
\hline
\multicolumn{3}{|c|}{BGL} \\ \hline 
\multicolumn{2}{|c|}{SDC} & $7.56^{+0.74}_{-0.43}$ \\ \hline
       0 &        0 & $7.60^{+0.74}_{-0.44}$ \\ \hline
       0 &    $\pi$ & $7.92^{+0.79}_{-0.46}$ \\ \hline
       0 & $3\pi/4$ & $7.77^{+0.75}_{-0.42}$ \\ \hline
 $\pi/2$ &    $\pi$ & $7.93^{+0.80}_{-0.46}$ \\ \hline
$3\pi/2$ &        0 & $7.60^{+0.74}_{-0.44}$ \\ \hline
\end{tabular}
\end{center}
\end{table}
\begin{table}[tb]
\begin{center}
\caption{\label{tab:taudeltaCBCL}
The total branching fraction for the $B^+ \to \pi^+ \tau^+ \tau^-$ decay in the BCL parametrization 
including the LD contribution from the $\psi(2S)$-meson for the various assumed values 
of the amplitude phases. SDC means the short-distance (perturbative) contribution.
}
\begin{tabular}{|c|c|c|}
\hline
$\delta^{(u)}_{\psi (2S)}$ & $\delta^{(c)}_{\psi (2S)}$ & ${\rm Br} (B^+ \to \pi^+ \tau^+ \tau^-) \times 10^{-9}$ \\
\hline
\multicolumn{3}{|c|}{BCL} \\ \hline 
\multicolumn{2}{|c|}{SDC} & $6.00^{+0.81}_{-0.49}$ \\ \hline
       0 &        0 & $5.79^{+0.78}_{-0.48}$ \\ \hline
       0 &    $\pi$ & $6.23^{+0.84}_{-0.50}$ \\ \hline
       0 & $3\pi/4$ & $6.05^{+0.80}_{-0.47}$ \\ \hline
 $\pi/2$ &    $\pi$ & $6.24^{+0.84}_{-0.51}$ \\ \hline
$3\pi/2$ &        0 & $5.78^{+0.78}_{-0.48}$ \\ \hline
\end{tabular}
\end{center}
\end{table}
\begin{table}[htb]
\begin{center}
\caption{\label{tab:taudeltaCmBCL}
The total branching fraction for the $B^+ \to \pi^+ \tau^+ \tau^-$ decay in the mBCL parametrization 
including the LD contribution from the $\psi(2S)$-meson for the various assumed values 
of the amplitude phases. SDC means the short-distance (perturbative) contribution.
}
\begin{tabular}{|c|c|c|}
\hline
$\delta^{(u)}_{\psi (2S)}$ & $\delta^{(c)}_{\psi (2S)}$ & ${\rm Br} (B^+ \to \pi^+ \tau^+ \tau^-) \times 10^{-9}$ \\
\hline
\multicolumn{3}{|c|}{mBCL} \\ \hline 
\multicolumn{2}{|c|}{SDC} & $6.28^{+0.76}_{-0.46}$ \\ \hline
       0 &        0 & $6.08^{+0.74}_{-0.46}$ \\ \hline
       0 &    $\pi$ & $6.50^{+0.80}_{-0.49}$ \\ \hline
       0 & $3\pi/4$ & $6.33^{+0.76}_{-0.45}$ \\ \hline
 $\pi/2$ &    $\pi$ & $6.51^{+0.80}_{-0.49}$ \\ \hline
$3\pi/2$ &        0 & $6.08^{+0.74}_{-0.44}$ \\ \hline
\end{tabular}
\end{center}
\end{table}

\subsection{The Ratios $R_\pi (\tau/\ell)$ ($ \ell =e,\mu$)}
\label{ssec:ratio-of-BFs}

\begin{figure}[htb]
\begin{center}
\includegraphics[width=0.35\textwidth]{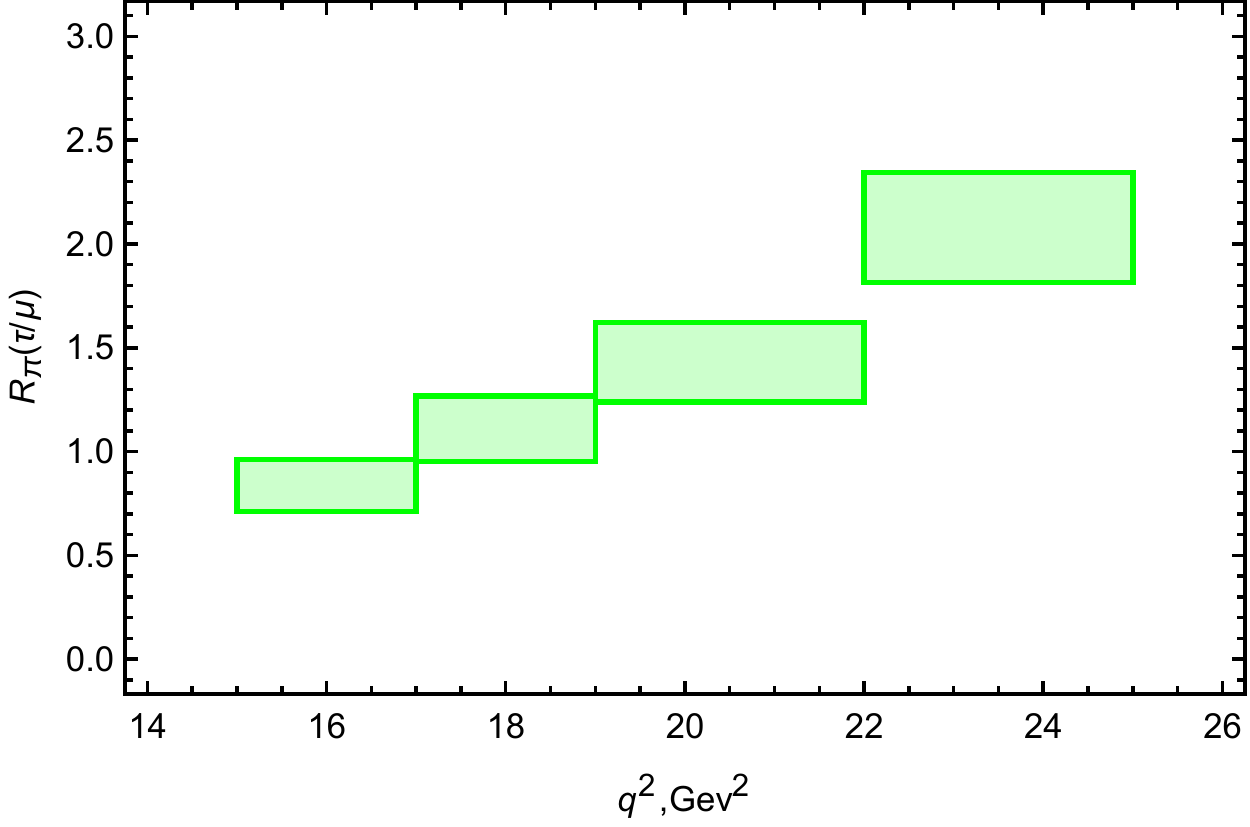} \\ 
\includegraphics[width=0.35\textwidth]{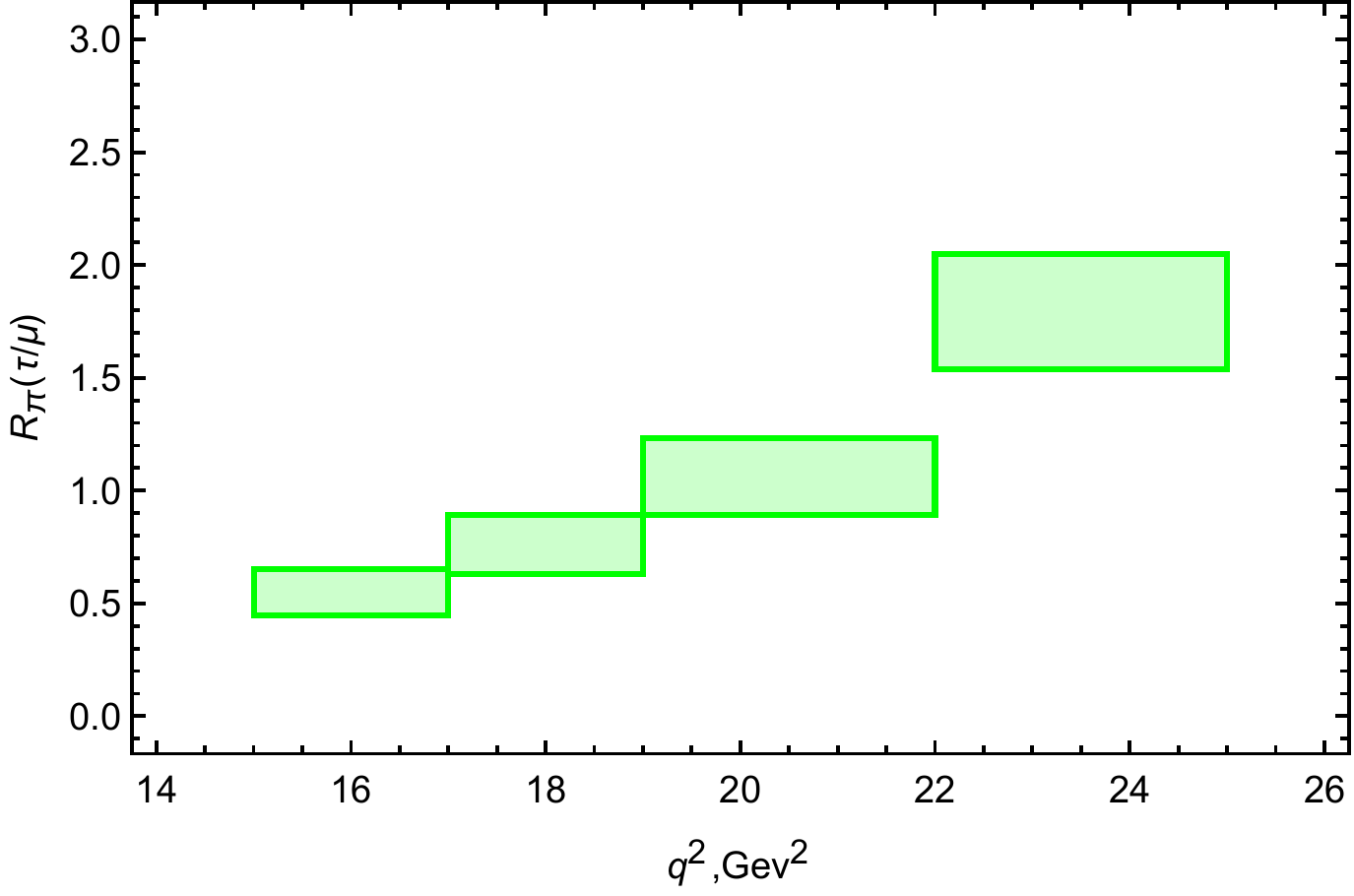} \\ 
\includegraphics[width=0.35\textwidth]{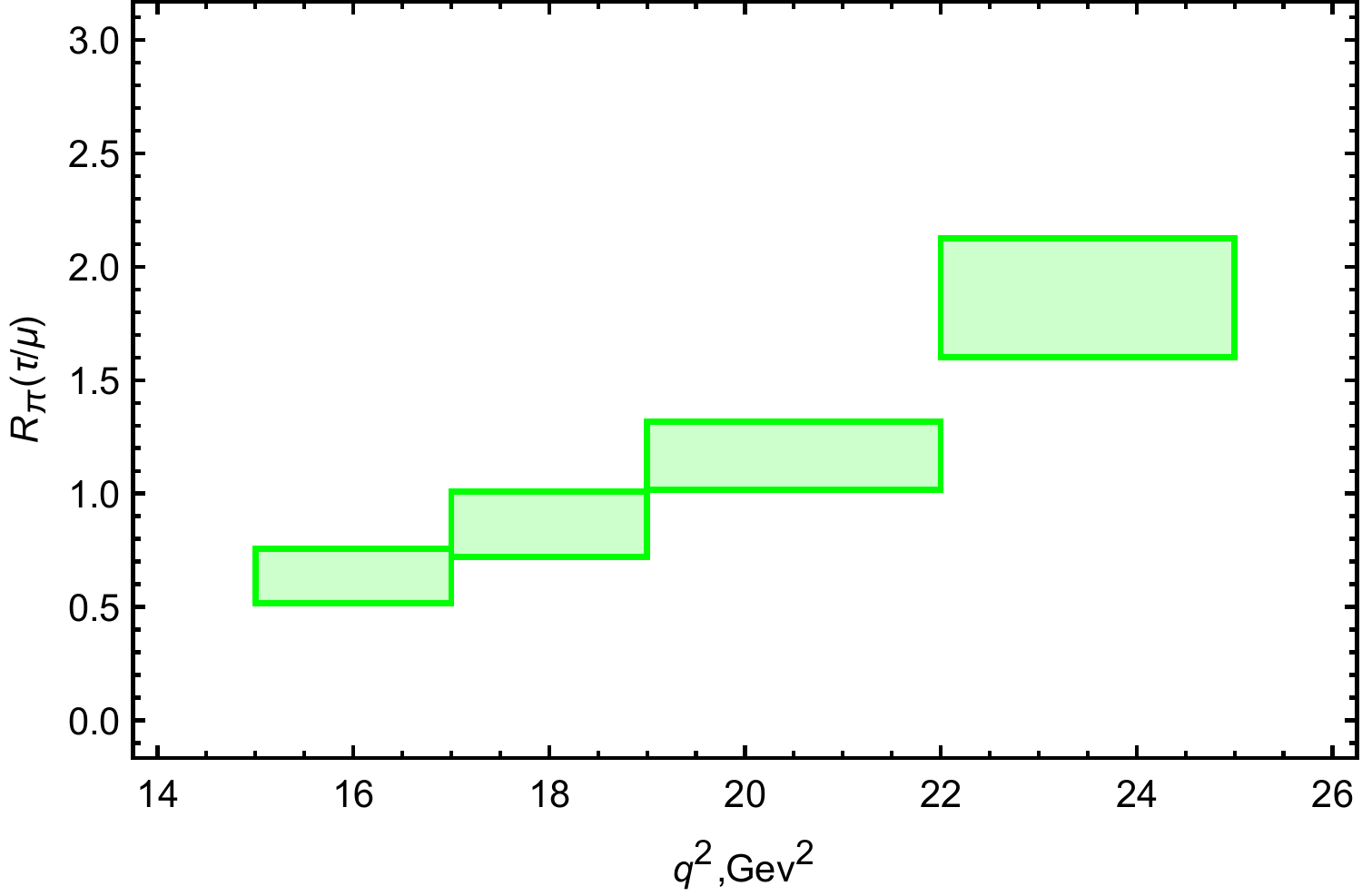}
\caption{\label{fig:binR}
Partial ratios $R_\pi^{(\tau/\mu)} (q^2_{\rm min}, q^2_{\rm max})$ in bins of the ditauon invariant 
mass squared for the BGL (top), BCL (center) and mBCL (bottom) form factor parametrizations.
}
\end{center}
\end{figure}

Since in the SM $R_\pi (\tau/\mu) = R_\pi (\tau/e)$ holds to a very high accuracy, we show 
numerical results only for $R_\pi (\tau/\mu)$, the ratio of the partially-integrated tauonic 
branching fraction $(\Delta{\rm Br})_\pi^\tau (q_1^2, q_2^2)$   % ${\rm Br} (B^+ \to \pi^+ \tau^+ \tau^-)$ 
to the muonic one $(\Delta{\rm Br})_\pi^\mu (q_1^2, q_2^2)$.    % ${\rm Br} (B^+ \to \pi^+ \mu^+ \mu^-)$.
The partial ratio is defined as follows:
\begin{equation}
R_\pi^{(\tau/\mu)} (q_1^2, q_2^2) \equiv
\frac{(\Delta{\rm Br})_\pi^\tau (q_1^2, q_2^2)}
     {(\Delta{\rm Br})_\pi^\mu  (q_1^2, q_2^2)} . 
\label{eq:R-pi-tau/mu-def} 
\end{equation}
To study the dependence of theoretical results on the choice of the FF parametrization, 
we plot the partial ratio $R_\pi^{(\tau/\mu)} (q^2_{\rm min}, q^2_{\rm max})$ in bins 
of the dilepton invariant mass squared, shown in Fig.~\ref{fig:binR}. Numerical values 
of this ratio are shown in Table~\ref{tab:comp_Rpart}, obtained by integrating the 
partial branching ratio over the indicated $q^2$-ranges. The errors shown take into account 
the uncertainties due to the factorization-scale, the CKM matrix element~$V_{td}$, and 
the form factor errors. Theoretical predictions for the total ratio for the BGL, BCL and 
mBCL parametrizations are as follows:
\begin{eqnarray}
\hspace{5mm} && R_\pi^{\rm BGL} (\tau/\mu)  = 0.44 \pm 0.16, 
\label{eq:R-pi-tau/mu-BGL} \\
\hspace{5mm} && R_\pi^{\rm BCL} (\tau/\mu)  = 0.31 \pm 0.12, 
\label{eq:R-pi-tau/mu-BCL} \\
\hspace{5mm} && R_\pi^{\rm mBCL} (\tau/\mu) = 0.37 \pm 0.15. 
\label{eq:R-pi-tau/mu-mBCL} 
\end{eqnarray} 
They agree with each other within the indicated uncertainties. The central values for $R_\pi (\tau/\mu)$ 
lie in the range $0.30 - 0.45$. The main uncertainty on $R_\pi (\tau/\mu)$, as opposed to the very 
precise ratio $R_{K^{(*)}} (\mu/e)$, is due to the form factors. These results are potentially 
useful in testing the lepton flavor universality in the $ b \to d \ell^+ \ell^-$ sector.

\begin{table}[tb]
\caption{\label{tab:comp_Rpart}
Theoretical predictions for the partial ratios 
$R_\pi^{(\tau/\mu)} ( q^2_{\rm min}, q^2_{\rm max})$. 
The boundaries of the $q^2$-intervals are in units of GeV$^2$.    
}
\begin{center}
{\small
\hspace{-3mm}
\begin{tabular}{|c|c|c|c|}
     \hline
 & \multicolumn{3}{c|}{$R_\pi^{(\tau/\mu)} ( q^2_{\rm min}, q^2_{\rm max})$} \\
\hline
[$q^2_{\rm min}, q^2_{\rm max}$] & BGL & BCL & mBCL \\
\hline
$[15.0 , 17.0]$ & $0.84 \pm 0.13$ & $0.55 \pm 0.10$ & $0.64 \pm 0.12$ \\
\hline
$[17.0 , 19.0]$ & $1.11 \pm 0.16$ & $0.76 \pm 0.13$ & $0.86 \pm 0.15$ \\
\hline
$[19.0 , 22.0]$ & $1.43 \pm 0.19$ & $1.06 \pm 0.17$ & $1.17 \pm 0.18$ \\
\hline 
$[22.0 , 25.0]$ & $2.08 \pm 0.27$ & $1.79 \pm 0.25$ & $1.86 \pm 0.26$ \\
\hline
\end{tabular}
}
\end{center}
\end{table}

\section{Summary and outlook}
\label{sec:summary}

We have presented theoretical predictions for the branching ratio ${\rm Br} (B^+ \to \pi^+ \tau^+ \tau^-)$ 
and the ditauon invariant-mass distribution at NLO accuracy in the SM, 
using three popular parametrizations of the $B \to \pi$ form factors, known in the literature 
as the BGL~\cite{Boyd:1995sq}, BCL~\cite{Bourrely:2008za}, and mBCL~\cite{Leljak:2021vte}. 
In the SM, LFU holds, which relates the decay $B^+ \to \pi^+ \tau^+ \tau^-$ to the observed 
decay $B^+ \to \pi^+ \mu^+ \mu^-$. In extensions of the SM, the LFU hypothesis can be easily 
violated, of which the leptoquark models are the primary candidates~\cite{Becirevic:2018afm,Crivellin:2022mff}.
In view of this, we have worked out the ratio of the branching ratios $R_\pi (\tau/\mu)$. 
Together with the corresponding ratios $R_\pi (\mu/e)$ and $R_\pi (\tau/e)$, their measurement 
will provide a precision test of the lepton flavor universality in the FCNC $b \to d$ transitions. 
Suffice to say that none of these ratios have been subjected to experimental scrutiny so far. 
Of these, the ratio $R_\pi (\mu/e)$ has been theoretically investigated in~\cite{Bordone:2021olx}. 
We have concentrated here on the ratios $R_\pi (\tau/\mu)$ and $R_\pi (\tau/e)$.  

The decay $B^+ \to \pi^+ \tau^+ \tau^-$ involves all three form factors, $f_+(q^2)$, $f_0(q^2)$, 
and $f_T(q^2)$. The uncertainties in ${\rm Br} (B^+ \to \pi^+ \tau^+ \tau^-)$ arise from the FF 
parametrizations, scale-dependence of the Wilson coefficients, and the CKM matrix element. 
Numerical values of the  SD (perturbative) contribution to ${\rm Br} (B^+ \to \pi^+ \tau^+ \tau^-)$ 
are listed in Table~\ref{tab:Br}. Partial branching fractions of the $B^+ \to \pi^+ \tau^+ \tau^-$ decay, 
$(\Delta{\rm Br})_\pi^\tau (q_{\rm min}^2, q_{\rm max}^2)$, in bins of ditauon 
invariant mass squared, are shown in Fig.~\ref{fig:bin} and displayed in Table~\ref{tab:tau}.

In addition, the LD-contribution from the process $B \to \pi V \to \pi \ell^+ \ell^-$ 
is calculated. Experimental data on the masses, partial and total decay widths of the charmonium states
are given in Table~\ref{tab:vector-charmonia}. Due to the small decay width of the $\psi(2S)$-meson, 
its LD-contribution is concentrated close to the $B^+ \to \pi^+ \tau^+ \tau^-$ reaction threshold and 
can be largely eliminated by a cut on the ditauon invariant mass squared ($q^2 \ge 15$~GeV$^2$) 
(see Fig.~\ref{fig:tauwres}). We have given arguments why the contribution from the higher charmonium 
resonances are numerically small, and hence they do not change the SD-contribution in this region perceptibly. 
Taking into account the LD contribution from the $\psi(2S)$-resonance, numerical results for the branching 
ratio ${\rm Br} (B^+ \to \pi^+ \tau^+ \tau^-)$ are given in Table~\ref{tab:taudeltaCBGL} for the BGL form 
factors. Various entries in this table correspond to using the indicated strong phases. The corresponding 
results for the BCL form factors are given in Table~\ref{tab:taudeltaCBCL} and for the mBCL form factors 
in Table~\ref{tab:taudeltaCmBCL}. Since the BGL parametrization and the Lattice-QCD based estimates 
are rather close to each other, our estimate of the total branching fraction is 
${\rm Br}_{\rm th} (B^+ \to \pi^+ \tau^+ \tau^-) =  7.5 \times 10^{-9}$, with an uncertainty of about 10\%. 

Numerical values for the ratio $R_\pi (\tau/\mu)$ are given in Eqs.~(\ref{eq:R-pi-tau/mu-BGL})--(\ref{eq:R-pi-tau/mu-mBCL}) 
for three parametrizations chosen. The central values lie in the range $0.30 - 0.45$. 
The main uncertainty on  $R_\pi (\tau/\mu)$, as opposed to the very precise ratio $R_{K^{(*)}} (e/\mu)$, 
is due to the form factors. Partial ratios $R_\pi^{(\tau/\mu)} (q^2_{\rm min}, q^2_{\rm max})$ in bins 
of ditauon invariant mass squared are shown in Fig.~\ref{fig:binR}. These ratios can be more precisely 
calculated in the future by progress in Lattice QCD.

The branching ratio for $B^+ \to \pi^+ \tau^+ \tau^- $, integrated over the region $q^2 \ge 15$~GeV$^2$,  
has a factor of about~3 suppression compared with the $B^+ \to \pi^+ \mu^+ \mu^-$ total branching fraction, 
which has already been measured. Of course, one has to factor in the experimental efficiency of reconstructing 
the $\tau^+ \tau^-$ pair, but a precise measurement is still feasible for the anticipated integrated 
luminosities at Belle~II and LHCb experiments. Once sufficient data is collected, also the measurements 
of the various asymmetries, such as the isospin-asymmetry involving $B^0 \to \pi^0 \ell^+ \ell^-$ and 
$B^\pm \to \pi^\pm \tau^+\tau^-$, and CP-violating asymmetries involving $B^+ \to \pi^+ \tau^+\tau^-$ 
and $B^- \to \pi^- \tau^+ \tau^-$ become interesting, which should be looked at theoretically.

{\it Acknowledgments}. 
I.~P. acknowledges the financial support of the German-Russian Foundation G-RISC 
and the kind hospitality of Christoph Grojean and the theory group at DESY, Hamburg, 
during her stay in Hamburg in the autumn of 2021.
A.~P. and I.~P. are supported by RSF  %  the Russian Science Foundation  
(Project No.~22-22-00877, https://rscf.ru/en/project/22-22-00877/). 

\appendix
\section{Expansion Coefficients of the FF Parametrization}
\label{app:1}

Values of the expansion coefficients in the BGL FF parametrization are borrowed 
from~\cite{Ali:2013zfa}. They are obtained as truncated series up to $k_{\rm max} = 2$ 
and presented in Table~\ref{tab:Values-BGL}.

\begin{table}[tb]
\caption{\label{tab:Values-BGL}
Values of the expansion coefficients in the BGL $B \to \pi$ form factor parametrization.}
\begin{center}
{\footnotesize % \small
\begin{tabular}{|c|c|c|c|}
     \hline
 & $f_+ (q^2)$ & $f_0 (q^2)$ & $f_T (q^2)$ \\
\hline
$a_0$ & $0.0209 \pm 0.0004$ & $0.0201 \pm 0.0007$ & $0.0458 \pm 0.0027$ \\
\hline
$a_1$ & $-0.0306 \pm 0.0031$ & $-0.0394 \pm 0.0096$ & $-0.0234 \pm 0.0124$ \\
\hline
$a_2$ & $-0.0473 \pm 0.0189$ & $-0.0355 \pm 0.0556$ & $-0.2103 \pm 0.1052$ \\
\hline 
\end{tabular}
}
\end{center}
\end{table}

Values of the expansion coefficients in the BCL parametrization are calculated from the
Lattice-QCD analysis presented in~\cite{FermilabLattice:2015cdh,FermilabLattice:2015mwy}. 
They are obtained as truncated series up to $k_{\rm max} = 3$ and collected in Table~\ref{tab:Values-BCL}.

\begin{table}[tb]
\caption{\label{tab:Values-BCL}
Expansion coefficients in the BCL FF parametrization.}
\begin{center}
{\small
\begin{tabular}{|c|c|c|c|}  
\hline
      &    $f_+ (q^2)$    &    $f_0 (q^2)$    &    $f_T (q^2)$    \\ \hline
$b_0$ & $0.407 \pm 0.015$ & $0.507 \pm 0.022$ & $0.393 \pm 0.017$ \\ \hline
$b_1$ & $-0.65 \pm 0.16$  & $-1.77 \pm 0.18$  & $-0.65 \pm 0.23$  \\ \hline
$b_2$ & $-0.46 \pm 0.88$  &  $1.27 \pm 0.81$  & $-0.60 \pm 1.50$  \\ \hline
$b_3$ &   $0.4 \pm  1.3$  &   $4.2 \pm 1.4$   &   $0.1 \pm 2.8$   \\ \hline 
\end{tabular}
}
\end{center}
\end{table}

\begin{table}[tb]
\caption{\label{tab:Values-mBCL}
Expansion coefficients in the mBCL FF parametrization.
}
\begin{center}
{\small
\begin{tabular}{|c|c|c|c|}  
\hline
          &       $f_+ (q^2)$       &       $f_0 (q^2)$      &       $f_T (q^2)$       \\ \hline
$f_i (0)$ &    $0.235 \pm 0.019$    &   $0.235 \pm 0.019$    &    $0.235 \pm 0.017$    \\ \hline
    $b_1$ & $-2.45^{+0.49}_{-0.54}$ & $0.40^{+0.18}_{-0.20}$ & $-2.45^{+0.45}_{-0.50}$ \\ \hline
    $b_2$ &   $-0.2^{+1.1}_{-1.2}$  &   $0.1^{+1.1}_{-1.2}$  & $-1.08^{+0.68}_{-0.71}$ \\ \hline
    $b_3$ &   $-0.9^{+4.2}_{-4.0}$  &      $3.7 \pm 1.6$     &    $2.6^{+2.1}_{-2.0}$  \\ \hline 
    $b_4$ &                         &     $1^{+14}_{-13}$    &                         \\ \hline 
\end{tabular}
}
\end{center}
\end{table}

Values of the expansion coefficients in the mBCL parametrization are obtained 
from the joint Light-Cone Sum Rules and Lattice QCD data~\cite{Leljak:2021vte}.
They are presented in Table~\ref{tab:Values-mBCL}.


\end{document}